\documentclass[aps,prd,twocolumn,nofootinbib,superscriptaddress]{revtex4}
\usepackage[hypertex]{hyperref}
\usepackage{graphicx}
\newcommand{\newc}{\newcommand}
\newc{\beq}{\begin{equation}}
\newc{\eeq}{\end{equation}}
\newc{\beqa}{\begin{eqnarray}}
\newc{\eeqa}{\end{eqnarray}}
\newc{\IM}{\mbox{\sl{Im}}}
\newc{\RE}{\mbox{\sl{Re}}}
\newc{\bit}{\begin{itemize}}
\newc{\eit}{\end{itemize}}
\newc{\CC}{\mathbf{c}}
\newc{\nonr}{\nonumber}
\newc{\hs}{\hskip 3mm}
\newc{\ra}{\rightarrow}
\newc{\TR}{\mbox{\sl{Tr}}}
\newc{\tri}{\triangle}
\newc{\xw}{\sin^2\theta_{\rm W}}
\newc{\trip}{$\mathbf{3}$ }
\newc{\sext}{$\bar{\mathbf{6}}$ }
\newc{\tripp}{$\mathbf{3^{\prime}}$ }
\begin{document}

% Page numbers bottom-center
\pagestyle{plain}

\title{
Phenomenology of a 5D Orbifold $SU(3)_W$ Unification model }

\author{We-Fu Chang}
\email{wfchang@triumf.ca}

\author{John N. Ng}
\email{misery@triumf.ca}
\affiliation{TRIUMF Theory Group,
4004 Wesbrook Mall, Vancouver, B.C. V6T 2A3, Canada}
\date{\today}

\begin{abstract}
We study the phenomenology of a  5D $SU(3)_W$ model on a $S_1/(Z_2\times Z'_2)$
orbifold
in which the minimal scalar sector plays an essential role of radiatively
generating neutrino Majorana masses without the benefits of  right-handed singlets.
We carefully examine  how do the exotic scalars
affect  the renormalization group (RG) equations for the gauge couplings
and the 5D $SU(3)_W$ unification.
We found that the compactification scale  of extra dimension is in the range of
$1/R\sim 1.5-5$ TeV. The possibility of the existence of
relatively low mass Kaluza-Klein
excitations makes the phenomenology of near term interest.
Some possible bilepton signatures can be searched for  in future
colliders and in neutrino scattering experiments with intense neutrino beams.
The low energy constraints from muon physics and lepton number violating
decay process induced by  bilepton are also discussed.
 These constraints can provide new information on the structure of Yukawa
couplings which might be useful for future model building.
\end{abstract} \pacs{Who cares?} \maketitle

\section{Introduction}
In 1971, an electroweak that unifies the $SU(2)\times U(1)$ symmetry of the
standard model(SM) into  $SU(3)_W$  was suggested \cite{Weinberg:1971nd}.
Here each family of the SM leptons is  grouped to form a $SU(3)_W$ fundamental
representation which is akin to the forgotten  Konopinski-Mahmoud \cite{Konopinski:1953gq}
assignments. The model predicts  the weak mixing angle to be $\xw=0.25$ at the tree
level. This is tantalizingly close to the measured value of $\xw=0.2311$.
Although the prediction is appealing, it was soon abandoned
because  the insurmountable theoretical difficulty to embed the quarks into
any $SU(3)_W$ multiplet.
Another relevant attempt  suggested  by
\cite{331model} is the so called 3-3-1 model.
In this model, the gauge group is  extended to $SU(3)_c\times SU(3)_W\times
U(1)$ such that the quarks can be embedded into it. This unification came with a hefty price.
Firstly, three extra exotic quarks were introduced
to complete the $SU(3)_W$ multiplets. Secondly, the scalar potential which
consists of three triplets and one sextet had to
be fine tuned to get the symmetry breaking
pattern right. Finally, the third family quarks had to be put
in a  representation which is different from the first two generations.

Recently, the development of orbifold grand unified theory(GUT) models in
the extra dimension brane world scenario has opened
up a new direction for model building. The old $SU(3)_W$ idea was
revived by several groups \cite{orbifold_SU3} by promoting it to a five dimensional (5D) model.
The $SU(3)_W$ is a gauge symmetry in the full 5D bulk space.
It is broken to the SM electroweak $SU(2)_L\times U(1)_Y$
by suitable choice of  orbifold parities. Then the  $SU(2)_L\times U(1)_Y$
symmetry is further
reduced  to $U(1)_{\rm EM}$  via the  usual Higgs mechanism.
In such a construction, the tree-level prediction  of $\xw=1/4$ is preserved and
the quarks can be placed at the orbifold fixed point where only the SM symmetry manifests; thus
avoiding  the long standing quark embedding problem. Clearly a dichotomy between quarks
and leptons still exists but it is no longer inconsistent.
A closer look at the Yukawa sector requires
at least one bulk scalar in ${\bf 3}$ representation  and one more bulk scalar
in the symmetrical ${\bf \bar{6}}$ representation
to give viable charged lepton mass pattern.
As a result, the electroweak symmetry breaking pattern is
controlled by the vacuum expected values(VEVs) of these two bulk scalars.

Meanwhile there are new developments in the area of experimental neutrino physics.
There is now strong evidence for  neutrino oscillations and nonzero neutrino masses
provided recently by many experiments \cite{SuperK, SNO, KamL, WMAP}
which  clearly  demands new physics beyond the SM for their explanation.
The most popular suggestion for giving the active  neutrino masses is the seesaw mechanism
in GUTs in which one right-handed SM singlet fermion with a mass
around GUT scale per SM family is introduced. Each right-handed
singlet  either belong to the fundamental representation, like
in $SO(10)$ GUT, or be put in by hand as in $SU(5)$ GUT. Usually, extra family
symmetry is required to obtain  neutrino masses and
mixing angles that are in accord with the known experimental measurements.
In extra dimensional models with right-handed bulk singlet(s), the
small four dimensional ( 4D ) effective Dirac masses can be a natural outcome of large extra dimension
volume dilution effect. Compared to bulk neutrino model, in a  4D
model, the neutrino masses can still emerge by incorporating  the  right-handed
singlet(s) but extreme fine tuning is required.
Both of above mentioned methods require a right-handed singlet going into
action.
An alternative way to give light neutrino masses is through 1-loop
quantum corrections. The prototype model was constructed some time
ago\cite{Zee} in which the presence of a $SU(2)$ singlet as the
lepton number violating source is crucial. The early version of
this kind of model gives at most the bi-maximal mixing angle
which is ruled out by the latest SNO data. At the expense of introducing
more exotic Higgs fields this
kind of models can be made to agree with observation again.
However, all these constructions suffer from being ad hoc.

Simultaneously confronting  $\sin^2\theta_W$ problem and recent neutrino data,
we  investigated a radiative mechanism  for neutrino mass based on 5D orbifold
$SU(3)_W$ GUT \cite{SU3:triumf}.
We found that it is possible to accommodate
large mixing angle MSW solution to the neutrino data with a minimum  scalar
sector, in the sense that  the charged lepton mass
hierarchy can be  accounted for, without much fine tuning of parameters.
This idea is further pursued  by us and has been extended to 5D non-supersymmetric $SU(5)$
model \cite{SU5:TRIUMF}. But due to the high compactification
scale,  $1/R>10^{14}$ GeV,  the $SU(5)$ version  is phenomenologically less interesting .
The non-supersymmetric $SU(3)_W$ model on the other hand implies a large extra
dimension with the compactification scale, $1/R$, in the range of $1.5-5$
TeV. Thus, the stability problem of the Higgs sector is not as severe as
many other theories. This point also justifies why the supersymmetry need not be
introduced in this model.
The theoretical  consistency, see the discussion in section III, sets a upper bound
on $1/R$ to be around $5$ TeV with the actual value  depending weakly on a few free
parameters.
Obviously, the low value of $1/R$ makes it interesting for colliders and
low energy physics tests of the validity of the model are meaningful.
Interestingly, all the new physics effects
appear mainly in the lepton sectors which makes it distinguishable from other GUT models.
For a brief review of these models see \cite{JR}.

The purpose of this paper is to carefully consider the
phenomenological consequences of the 5D $SU(3)_W$ GUT.
Nontrivial result already can be drawn from analyzing renormalization group equations (RGEs)
and unification altered by the extended scalar sector.

The characteristic phenomenology of $SU(3)_W$ GUT arises from the presence of
 vector or scalar bileptons.
There exists  many  phenomenology surveys on the general bilepton case
\cite{Cuypers, bilepton_gen}. Also there are studies that focused either on
the 3-3-1 model \cite{331pheno, Dion:1998pw}
or other GUTs \cite{bilepton_GUT}.
The presence of Kaluza-Klein ( KK ) excitations of bileptons or SM gauge
bosons in this 5D $SU(3)_W$ model
makes our study very different  from the existing analysis.
The KK excitations will affect the effective gauge and Yukawa couplings.
Signatures in  high energy collider experiments can come from
 direct production of new particles if they are light enough. They can also come from
modification of the expected SM signature.
Depending  on the mixings which can occur among the vector bileptons and also the leptons, we found
the deviation of angular distributions from SM expectations can be as high as few tens percent
in the Bhabha and M\o ller scatterings which could be probed by the future linear colliders.
Also, in low energy neutrino scattering the KK excitations play a role.

Current low energy leptonic experiment such as rare muon processes already  impose many
 constraints on
this model, especially on the Yukawa sector as we will show.

Here is our plan for the paper.
In the next section, we give necessary details of the 5D orbifold $SU(3)_W$
model. The coupling of the bilepton gauge bosons to leptons will
be spelled out explicitly. The RGEs for the gauge couplings are
examined in section III. This is done in
greater detail than available in the literature.
Then the tree-level decay widths of KK excitations are given in section IV.
The low energy constraint will be discussed in section V, where muon decay,
 muonium-antimuonium transition, and lepton flavor violating processes
are discussed. Section VI is devoted to the study of collider signatures,
focusing on the Bhabha and M\o ller scatterings. Also, we
comment on the possible bilepton direct production in various colliders.
The low energy neutrino electron scattering receives correction
from the KK excitations of SM gauge bosons and the vector bileptons. This
will be discussed in section VII.
We give the conclusion at section VIII.
Some handy formulas are  collected  in the Appendix.

\section{The 5D orbifold $SU(3)_W$ model }
The $SU(3)_W$ unification model is defined on the orbifold
$M_4\times S_1/(Z_2\times Z'_2)$ with the
4D Minkowski space $M_4$ coordinates denoted by $x^{\mu} (\mu =0,1,2,3)$ and the
extra spatial dimension by $y$. The latter is compactified into a
 circle  $S_1$ of  radius $R$, or $y=[-\pi R, \pi R]$, and is orbifolded by a $Z_2$
which identifies points $y$ and $-y$. The resulting space is further divided by a
second $Z_2^{\prime}$ acting  on $y'= y - \pi R/2$ to give the final geometry.
Therefore, two parity eigenvalues, plus or minus one, can be assigned to the bulk fields
under the transformations  $P: y\leftrightarrow -y$ and $P': y'\leftrightarrow -y'$.

The SM lepton left-handed doublet and the right-handed singlet in each family
can be embedded into the $SU(3)_W$ fundamental
representation as
\beq
L=\left(\begin{array}{c} e\\ \nu \\ e^c \end{array} \right)_{\mathrm L}.
\eeq
In our convention, $T_3=-\lambda_3/2$ and $Y= -\sqrt{3} \lambda_8/2$ in terms of the
standard Gell-Man matrices and ${\mathrm L}=(1-\gamma _5)/2$.
The $SU(3)_W$ gauge bosons are bulk fields and the 5D
field strength is given by
\beq
G_{MN}= \partial_M {\cal A}_N- \partial_N {\cal A}_M -{i\tilde{g} \over
\sqrt{M^*}}[{\cal A}_M, {\cal A}_N]
\eeq
with  gauge matrix ${\cal A}_M= \sum_{a=1}^8 A^a_M T^a$,
 the generator $T^{a}=\frac12 \lambda^{a}$ and  $M,N=\{\mu, y\}$.
It's more convenient to express the  gauge matrix ${\cal A}_M$ in gauge boson's mass
basis:
\beqa
 {\cal A}_M=
{1\over\sqrt{2}}\left(\begin{array}{ccc}
  0 & W^-_M & U^{-2}_M \\ W^+_M& 0& V^-_M\\ U^{+2}_M& V^+_M& 0 \end{array}
  \right)\nonr\\
+ { Z_M\over c_W}\left(\begin{array}{ccc}
  {c_W^2-s_W^2 \over 2} & 0 & 0 \\ 0& -\frac12 &0\\ 0&0& s_W^2 \end{array} \right)
+ A_M \left(\begin{array}{ccc}
  s_W & 0 & 0 \\ 0& 0&0\\ 0&0&-s_W \end{array} \right)
\eeqa
where
\beq
\left( \begin{array}{c}  Z_M \\ A_M \end{array}\right)
=\left( \begin{array}{cc}  c_W&-s_W \\ s_W &c_W \end{array}\right)
\left( \begin{array}{c}  A^3_M \\ A^8_M \end{array}\right)
\eeq
and  $\tilde{e}= \tilde{g}\sin\theta_W$.
The key to orbifold symmetry breaking is that the individual components of  a gauge
multiplet need not share the same $(P,P')$ parities.  It is only required
that the all the  assignments to bulk fields  are consistent with
group theory  and $P,P'$ which are inner automorphisms of $SU(3)_W$.
The parity matrices ${\rm P}=diag\{+++\}$ and ${\rm P}'=diag\{++-\}$
and the parity of $SU(3)_W$ gauge fields
\beqa
{\cal A}_\mu(y) = {\rm P} {\cal A}_\mu(-y) {\rm P}^{-1}\,,&&
{\cal A}_\mu(y') = {\rm P}' {\cal A}_5(-y') {\rm P}'^{-1}\nonr\\
{\cal A}_5 (y) = -{\rm P} {\cal A}_\mu(-y){\rm P}^{-1}\,,&&
{\cal A}_5(y') = {\rm P}' {\cal A}_5(-y'){\rm P}'^{-1}
\eeqa
are chosen to break the bulk $SU(3)_W$ symmetry into $SU(2)\times U(1)$.
Then the $(Z_2,Z'_2)$ parities of the SM gauge bosons and $SU(3)_W/(SU(2)\times U(1))$ gauge
bosons denoted by $U^{\pm2}, V^\pm$, are $(++)$ and $(+-)$ respectively.
In other words, only SM gauge bosons have zero modes. Both the $U,V$ gauge bosons
and all the $y-$components are heavy KK excitations. Note that the KK modes of the
fifth components of gauge bosons are real scalars.

The $SU(3)_W$ symmetry is explicitly broken to $SU(2)_L\times
U(1)$ at  the $y=\pi R/2$ fixed point, where
the  4D quarks field are forced to live on it.
On the other hand, the lepton fields can be placed anywhere in the bulk or on either
two fixed points.
We choose to put the 4D lepton triplets at $y=0$ which is a $SU(3)_W$ symmetric
fixed point so that they  enjoy the $SU(3)_W$ symmetry. This also avoids possible
proton decay contact interactions.

As mentioned before,  one Higgs triplet $\mathbf{3}$ plus one Higgs anti-sextet
\sext, denoted as $\phi_6$, is the minimal scalar set  to give viable charged fermion
masses (see \cite{SU3:triumf} ).
The parity of scalar fields are chosen to be
\beqa
\phi_3(y)= {\rm P} \phi_3(-y)\,,&& \phi_3(y')= {\rm P}' \phi_3(-y')\nonr\\
\phi_6(y)= {\rm P} \phi_6(-y){\rm P}^{-1}\,,&& \phi_6(y')=- {\rm P}' \phi_6(-y') {\rm P}'^{-1}.
\eeqa
This will  allow the appropriate neutral fields to develop  VEVs and also avoids tree-level
neutrino masses.
Another Higgs triplet \tripp with parities $(+-)$ is introduced
to transmit lepton number violation essential  for generating Majorana neutrino mass
through 1-loop diagrams \cite{SU3:triumf}.
This comes from  a triple Higgs interaction of  the type of
${\mathbf 3'}^T \bar{{\mathbf 6}}{\mathbf 3}$ which is consistent with
the symmetry of the background geometry.

Now we have all the ingredients to write down explicitly the  5D Lagrangian density
\beqa
{\cal L}_5 &=& -\frac12 \TR[G_{MN}G^{MN}]
 +\TR[(D_M\phi_6)^\dag(D^M\phi_6)]\nonr\\
&+& (D_M\phi_3)^\dag(D^M\phi_3)
 +(D_M\phi'_3)^\dag(D^M\phi'_3)\nonr\\
&+& \delta(y) \left[\epsilon_{abc}\frac{ f_{ij}^3}{\sqrt{M^*}} \overline{(L^a_i)^\CC} L^b_j \phi_3^c
 + \epsilon_{abc}\frac{f_{ij}^{'3}}{\sqrt{M^*}}  \overline{(L^a_i)^\CC} L^b_j \phi_3^{'c}
\right.\nonr\\
&+& \left. \frac{f^6_{ij}}{\sqrt{M^*}}\overline{(L^a_i)^\CC} \phi_{6 \{ab\}} L^b_j
 + \bar{L}i\gamma^\mu D_\mu L\right]\nonr\\
&-& V_0(\phi_6,\phi_3,\phi'_3) -
 \frac{m}{\sqrt{M^*}}\phi_3^T \phi_6 \phi'_3 + H.c.\nonr\\
&+& \mbox{quark sector}.
\label{5DL}
\eeqa
The 5D covariant derivatives are
given by
\beqa
D_M\phi_3= (\partial_M- i {\tilde{g}\over \sqrt{M^*}} {\cal A}_M )\phi_3,\\
D_M\phi_6= \partial_M \phi_6- i {\tilde{g}\over \sqrt{M^*}} [ {\cal A}_M\phi_6+
({\cal A}_M\phi_6 )^T ]\,.
\eeqa
The notations are self explanatory. The cutoff scale $M^*$ is
introduced to make the coupling constants  dimensionless.
The quark sector is not relevant now and will be left out.
The complicated scalar potential is gauge invariant and orbifold symmetric
and will not be specified since it is not needed here.
Since we will only concentrate on tree level processes in this paper,
 the gauge fixing term is not specified here. This can
be done either covariantly or non-covariantly \cite{GF}.

The branching rules of bulk fields and the labels
to each components  are summarized below:
\beqa
8^\mu &=& \underbrace{(1,0)_{++}}_{B^\mu}
+\underbrace{(3,0)_{++}}_{A^\mu}
+\underbrace{(2,-3/2)_{+-}  + (2,+3/2)_{+-}}_{(U,V)^\mu}\nonr\\
8^5 &=& \underbrace{(1,0)_{-+}}_{B^5}
+\underbrace{(3,0)_{-+}}_{A^5}
+\underbrace{(2,-3/2)_{--}  + (2,+3/2)_{--}}_{(U,V)^5}\nonr\\
\mathbf{3} &=& \underbrace{(2,-1/2)_{++}}_{H_{W1}}
+\underbrace{(1,1)_{+-}}_{H_{S}}\nonr\\
\mathbf{3}' &=& \underbrace{(2,-1/2)_{+-}}_{H'_{W1}}
+\underbrace{(1,1)_{++}}_{H'_{S}}\nonr\\
\mathbf{\bar{6}} &=& \underbrace{(3,+1)_{+-}}_{H_{T}}
+\underbrace{(2,-1/2)_{++}}_{H_{W2}}+\underbrace{(1,-2)_{+-}}_{H_{S2}}\nonr
\eeqa
where the SM  quantum numbers are $(SU(2)_L,U(1)_Y)$ and the subscripts
label the parities $P, P^{\prime}$.
Then it is straightforward to obtain the 4D effective interaction by
integrating over $y$.
In the following we list some relevant Lagrangian for our phenomenological study.

The 4D effective gauge coupling can be identified as $g_2= \frac{\tilde{g}}{\sqrt{2\pi R M^*}}$
which gives the following  charged current interaction
\beqa
{\cal L}_{\rm CC}=  {g_2 \over \sqrt{2}}\sum_{n=0}\kappa_n  \left[
\overline{e_L}\gamma^\mu \nu_L W^-_{n,\mu}
 +\overline{e_L}\gamma_\mu e_R^\CC U^{-2}_{n,\mu}\right.\nonr\\
\left.+\overline{\nu_L}\gamma^\mu e^\CC_R V^{-1}_{n,\mu}
+H.c.
\right]
\eeqa
where $\kappa_n=(\sqrt{2})^{1-\delta_{n,0}}$.
Similarly, the neutral current interactions are  worked out to be
\beqa
{\cal L}_{\rm NC}=  \sum_{n=0}\kappa_n {g_2\over \cos\theta}
  \bar{l}\gamma_\mu(g^l_L P_{L}+g^l_R P_{R})l Z^\mu_n\nonr\\
  - \sum_{n=0} \kappa_n e (\bar{l}\gamma_\mu l) A^\mu_n
  +  \sum_{n=0} \kappa_n {g_2\over 2\cos\theta}
  (\bar{\nu}\gamma_\mu P_{L}\nu) Z^\mu_n\,.
\eeqa
The usual SM relations hold: $e=g_2 \sin\theta_W$,
$g^l_{L/R}= T_3(l_{L/R})-\xw Q(l_{L/R})$,
and there is no extra neutral current except for the KK modes of the  photon and the Z boson.
The explicit expression of the Yukawa interactions is given by
\begin{widetext}
\beqa
{\cal L}_{\rm Y}
&=& { f^3_{ij}\over \sqrt{2\pi R M^*}}\sum_{n=0} \kappa_n  \left [
\overline{e^\CC_{L,i}}\nu_{L,j} H^+_{S,n}
+ \overline{e_{R,i}}e_{L,j} H^0_{W1,n}
+ \overline{e_{R,j}}\nu_{L,i} H^-_{W1,n} -(i\Leftrightarrow j)
\right]\nonr\\
&+&
{f^{'3}_{ij}\over \sqrt{2\pi R M^*}}\sum_{n=0}\kappa_n  \left [
\overline{e^\CC_{L,i}}\nu_{L,j} H^{'+}_{S,n}
+ \overline{e_{R,i}}e_{L,j} H^{'0}_{W1,n}
+ \overline{e_{R,j}}\nu_{L,i} H^{'-}_{W1,n} -(i\Leftrightarrow j)
\right]\nonr\\
&+&{f^6_{ij}\over \sqrt{2\pi R M^*}}\sum_{n=0} \kappa_n  \left [
 \overline{e^\CC_{L,i}}e_{L,j} H^{+2}_{T,n}
+( \overline{e^\CC_{L,i}}\nu_{L,j}
+ \overline{\nu^\CC_{L,i}}e_{L,j} ) H^+_{T,n}
+ \overline{\nu^\CC_{L,i}}\nu_{L,j} H^0_{T,n} \right.\nonr\\
&+&\left. ( \overline{e_{R,i}}e_{L,j} + \overline{e_{L,i}}e_{R,j}) H^0_{W2,n}
+ (\overline{e_{R,i}}\nu_{L,j}+ \overline{\nu_{L,i}}e_{R,j}
)H^-_{W2,n} +  \overline{e_{R,i}} e^\CC_{R,j} H^{-2}_{S2,n}
\right] + H.c.
\eeqa
\end{widetext}
After spontaneously symmetry breaking, the \trip  and the \sext
acquire non-zero VEVs:
\beq
<\mathbf{3}> = \frac{v_3^{3/2}}{\sqrt{2}}\left( \begin{array}{c} 0 \\1\\0
\end{array}\right),
<\mathbf{\bar{6}}> = \frac{v_6^{3/2}}{\sqrt{2}}\left( \begin{array}{ccc}
 0&0&{1\over \sqrt{2}} \\ 0&0&0\\{1\over \sqrt{2}}&0&0
\end{array}\right)
\eeq
which are  related to the 4D VEV as
\beq
v_0^2= 2\pi R (v_6^3+v_3^3 )\sim (250 \mbox{GeV})^2
\eeq
The gauge boson masses can be derived to be:
\beqa
M_{Wn}^2&=& M_{W0}^2+ \frac{4n^2}{R^2},\\
M_{Zn}^2&=&M_{Z0}^2+\frac{4n^2}{R^2},\\
M_{Vn}^2&=& M_{W0}^2+ \frac{(2n-1)^2}{R^2},\\
M_{Un}^2&=& {4M_{W0}^2 \over 1+v_3^3/ v_6^3
}+\frac{(2n-1)^2}{R^2}\, .
\eeqa
Unless we specify otherwise, in later numerical estimations we shall ignore the zero
mode masses for KK excitations.
The two VEVs, $v_3$ and $v_6$, are free parameters. There is no good reason for them
to be very different. For the sake of economy
 we assume $v_3=v_6$; this also implies that all
flavor information are encoded in the Yukawa couplings.

\begin{figure}[tc]
\centering \includegraphics[width=0.9\columnwidth]{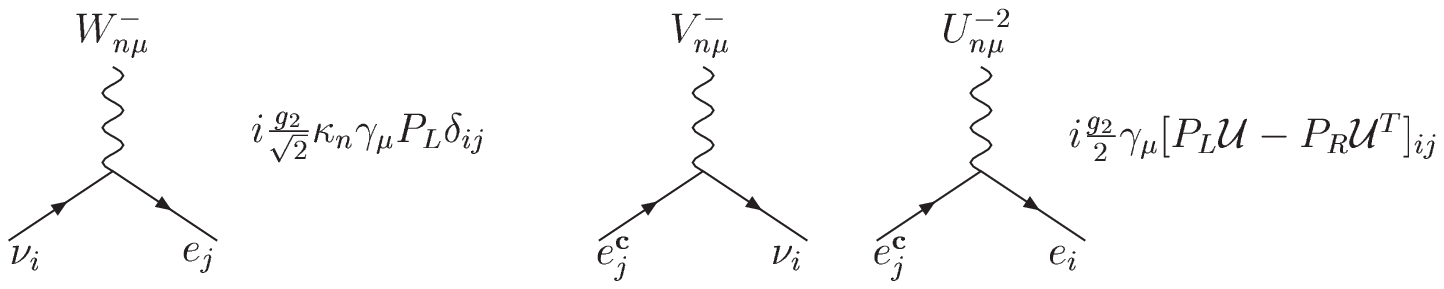}
\caption{Feynman rules for charged current vertices.}
\label{fig:Feyrule_UV}
\end{figure}

Denoting  $l=(e,\mu,\tau)^T$, the charged leptons acquire masses through VEVs as usual
with
\beq
 (\overline{l_R})^T{\cal M}_l l_L  + H.c.
\eeq
and the mass matrix is given by
\beqa
{\cal M}_l= {v_0 \over \sqrt{2\pi R M^*}} \left(\begin{array}{ccc}
  0& f^3_{12} & f^3_{13} \\ -f^3_{12}& 0 &f^3_{23} \\ -f^3_{13}& -f^3_{23}&0 \end{array} \right)
\nonr\\
+{v_0 \over 2\sqrt{\pi R M^*}} \left(\begin{array}{ccc}
 f^6_{11}& f^6_{12} & f^6_{13} \\ f^6_{12}&  f^6_{22} & f^6_{23} \\ f^6_{13}& f^6_{23}& f^6_{33} \end{array}
 \right)\,.
\eeqa
The above mass matrix can be diagonalized by a bi-unitary rotation
with $\psi_{L/R}=U_{L/R} \psi'_{L/R}$ where the fields with prime are
weak eigenbasis. In this convention,
\beq
U_R {\cal M}_l {\cal M}_l^\dag U_R^\dag =
U_L {\cal M}^\dag_l {\cal M}_l U_L^\dag ={\cal M}^2_{diag}\,.
\eeq
It is straightforward to modify Feynman rules by incorporating the
mixing matrices. For example, the lagrangian of charged current in
charged leptons' mass eigenstate becomes:
\beqa
{\cal L}_{\rm CC} = g_2 \sum_{n=1}
\overline{e_{L i}}\gamma^\mu P_L (U_L^\dag U_R^*)_{ij}
e^\CC_{R j} U^{-2}_{n,\mu} +H.c.\nonr\\
+ g_2 \sum_{n=1}
\overline{\nu_{L i}}\gamma^\mu P_L (U_L^\dag U_R^*)_{ij}
e^\CC_{Rj} V^{-1}_{n,\mu} +H.c.
\eeqa
where the subscripts $L$ and $R$ are kept for book keeping.
For later use, we define the mixing matrix
\beq
{\cal U}=U_L^\dag U_R^*
\eeq
which is a CKM-like unitary matrix.
By using the identities in the Appendix, the above expression
can be rewritten as
\beqa
{\cal L}_{\rm CC} &=& {g_2\over 2} \sum_{n=1}
\overline{e_{i}}\gamma^\mu [ P_L {\cal U}- P_R {\cal U}^T]_{ij}
e^\CC_j U^{-2}_{n,\mu} +H.c. \nonr\\
&+& \frac{g_2}{2} \sum_{n=1}
\overline{\nu_{i}}\gamma^\mu[  P_L {\cal U}- P_R {\cal U}^T]_{ij}
e^\CC_j V^{-1}_{n,\mu} +H.c.
\eeqa
The corresponding Feynman diagrams are shown in Fig.\ref{fig:Feyrule_UV}, note that
the $W$ vertex is always flavor diagonal.

For completeness, we present the resulting neutrino overall mass scale.
This is calculated to be \cite{SU3:triumf}
\beq
\label{eq:numass}
\bar{m}_\nu \sim {g v_0 \over 16 \pi^2 M_W}
{m |f^{'3}| \over (\pi R M^*)}{m_\tau^2 \ln(M_2^2/M_1^2) \over M_1^2-M_2^2}
\eeq
where $M_1$ and $M_2$ are the masses of $H_{W2}$ in $\mathbf{\bar{6}}$ and
$H'_S$ in $\mathbf{3}'$ for the leading contribution.

\section{The RG running of coupling constants}
Below the scale $1/R$ the RG running of the gauge couplings are the same as the 4D
field theory.
When $\mu\gg 1/R$, the inverse fine structure constants obey  the following equation:
\beqa
{1\over \alpha_i(\mu)}={1\over \alpha_i(M_Z)} -{a_i\over 2\pi
}\ln{\mu \over M_Z}  - {a^H_i\over 2\pi }\ln{\mu\over M_H }\nonr\\
- {\tilde{a} \over 4\pi}(\mu R-\ln 2)+{\tilde{a}_i^e \over
4\pi}\ln\left({\pi\mu R  \over 2}\right)
\eeqa
where the beta function coefficients are denoted generically by $a$ and the subscript $i$
is for either $SU(2)$ or $U(1)$. Those with a tilde on top
are from  KK modes. Also
$\tilde{a}\equiv \tilde{a}^e+\tilde{a}^o$
 where $\tilde{a}_i^{e,o}$ is the beta function coefficient from the {\it
even(odd)}-KK mode. By {\it even(odd)} we refer to the component fields
with $(Z_2,Z'_2)$ parity $(++)$ or $(--)$ (\ $(+-)$ or $(-+)$\ )
and tree level KK masses of $2n/R$ (\ $(2n-1)/R$\ ).
The well known power law running was first noticed by \cite{Dienes} which can
also be understood by summing up KK modes contribution level by
level and then applying the Stirling's approximation \cite{SU5:TRIUMF}.
Note that the coefficient in front of  the  power law running term,
$(\mu R-2)$, is universal for  every subgroup and plays no role in determining the point of
unification. This  is no surprise because when $\mu \gg 1/R$ all the acting particles can
be clustered into some kind of GUT multiplets. The effect of even- and odd-components
splitting in a GUT multiplet shows up in the last term. As unification is concerned,
$\tilde{a}^e_i$ can be equivalently replaced by $-\tilde{a}^o_i$.
The beta function coefficient $a_i$ is  determined by the well-known formula:
\beq
\label{eq:beta_coef}
a_i=\left[-\frac{11}{3}S^i_2(G)+\frac23 S^i_2(F)+\frac16 S_2^i(S)\right]
\eeq
obtained from gauge boson self-energy corrections. In Eq.(\ref{eq:beta_coef}) the first term
comes from the gauge boson loops (G); the second one is from  Weyl fermion (F) loops;
the last one is due to  real scalar (S) loops and
for complex scalars this  should be doubled.
$S_2$
are standard group theory factors and they depend on
 the group representations of the respective particles in the loop.
The  hypercharge coupling is normalized to the unified gauge coupling as
$g_Y=g_{GUT}/\sqrt{3}$, and  the tree level prediction of
$\xw=1/4$ follows immediately.
The contributions from  individual fields and their KK modes are listed
below in  Table \ref{tab:beta_a}.
\begin{table}[ht]\begin{center}
\begin{tabular}{ccccc}  \hline
Sources & Component & $a_1$ & $a_2$ & Whole multiplet \\ \hline \hline
$\mathbf{3}(+,\pm)$
&  $(2,-1/2)_{(+,\pm)}$ & $1/18$ & $1/6$ & $1/6$ \\
&  $(1,1)_{(+,\mp)}$    & $1/9$ & $0$ & \\
\hline
$\mathbf{\bar{6}}(+,-)$
&  $(2,-1/2)_{(++)}$ & $1/18$ & $1/6$ & $5/6$ \\
&  $(3,1)_{(+-)}$ & $1/3$ & $2/3$ &  \\
&  $(1,-2)_{(+-)}$    & $4/9$ & $0$ & \\
\hline
$8^\mu(++)$
&  $(3,0)_{(++)}$ & $0$ & $-22/3$ & $-11$ \\
&  $(1,0)_{(++)}$ & $0$ & $0$ & \\
&  $(2,-3/2)_{(+-)}$ & $-11/2$ & $-11/6$ & \\
&  $(2,+3/2)_{(+-)}$ & $-11/2$ & $-11/6$ & \\
\hline
$L^i$ & $(1,2,-1/2)_b$ & $1/9$ & $1/3$ & $(20/27, 4/3)$\\
$e_R^i $ & $(1,1,-1)_b$ & $2/9$ & $0$ & \\
$Q^i $ & $(3,2,1/6)_b$ & $1/27$ & $1$ & \\
$u_R^i $ & $(3,1,2/3)_b$ & $8/27$ & $0$ & \\
$d_R^i $ & $(3,1,-1/3)_b$ & $2/27$ & $0$ & \\
\hline
&&&&\\
KK & Even component & $\tilde{a}^e_1$ & $\tilde{a}^e_2$ & $\tilde{a}$ \\ \hline \hline
$\mathbf{3}(++)$
&  $(2,-1/2)$ & $1/18$ & $1/6$ & $1/6$ \\
\hline
$\mathbf{3}'(+-)$
&  $(1,1)$    & $1/9$ & $0$ & $1/6$ \\
\hline
$\mathbf{\bar{6}}(+-)$
&  $(2,-1/2)$ & $1/18$ & $1/6$ & $5/6$ \\
\hline
$8^\mu(++)$
&  $(3,0)$ & $0$ & $-22/3$ & $-11$ \\
&  $(1,0)$ & $0$ & $0$ & \\
\hline
$8^5(-+)$
&  $(2,-3/2)$ & $1/4$ & $1/12$ & $1/2$ \\
&  $(2,+3/2)$ & $1/4$ & $1/12$ & \\
\hline
\end{tabular}\end{center}
\caption{Beta function coefficients form varies particles.}
\label{tab:beta_a}\end{table}

The unification condition can be expressed as
\beqa
{1\over \alpha_1(M_Z)}-{1\over \alpha_2(M_Z)}=
{a_1-a_2\over 2\pi}\ln{\mu \over M_G}\nonr\\
+{a^{H_3}_1-a^{H_3}_2 \over 2\pi }\ln{M_{H_3} \over M_G}
+{a^{H_6}_1-a^{H_6}_2 \over 2\pi }\ln{M_{H_6} \over M_G}\nonr\\
-{\tilde{a}_1^e-\tilde{a}_2^e \over 4\pi}\ln\left({\pi R M_G \over 2}\right)
\eeqa
where $M_{H_3}$ and $M_{H_6}$ are the zero mode masses for
$\mathbf{3}'$ and $\mathbf{\bar{6}}$ respectively.
Plugging in the numbers, $a=(41/18, -19/6)$ $= 3\times(20/27,4/3)$
$ + (0,-22/3)+(1/18,1/6)$,
$a_H^3=(1/9,0)$, $a_H^6=(1/18,1/6)$ and $\tilde{a}^e=(13/18,-41/6)$,
we have
\beqa
{2\pi\over \alpha_1(M_Z)}-{2\pi\over \alpha_2(M_Z)}=
-{49\over 9}\ln{M_Z \over M_G}
+{1\over 9 }\ln{M_{H_6} \over M_{H_3}}\nonr\\
-{34 \over 9}\ln\left({\pi R M_G \over 2}\right).
\label{eq:unieq}
\eeqa
Note that the dependence
of $M_{H_3}$ and $M_{H_6}$ is very weak since they appear in a ratio
in the above although they are strictly unknown parameters of the model.

The values of the couplings at $M_Z$ are very accurately measured
and is given by  $\alpha^{-1}(M_Z)=127.934(7)$ and
$\sin^2\theta_W= 0.23113(15)$. Using
\beq
\alpha_1= {3\alpha \over \cos^2\theta_W}=0.03049(2),\,
\alpha_2={\alpha \over \sin^2\theta_W}=0.03382(2)\,.
\eeq
Eq.(\ref{eq:unieq}) can be further simplified numerically to
\beq
8.2398= \ln{M_G \over(\mbox{GeV})}+{1\over 49}\ln{M_{H_6} \over M_{H_3}}
-{34\over 49}\ln\left({\pi R M_G \over 2}\right)
\eeq
which can be  easily solved.
Some typical solutions are given in Tab.\ref{tab:RG_solutions}.
The compactified extra space is large and of order few TeV as advertised earlier.
The results  agrees with  previous estimations \cite{orbifold_SU3}. Since we
do not expect a large hierarchy in the scalar masses this is fairly robust.
Note that upper bound of $1/R$ is determined  by the requirement of theory
consistency, namely $M_G>1/R$.
\begin{table}[ht]
\begin{center}
\begin{tabular}{ccc}
$M_{H_6}/M_{H_3}$ & $(\pi R M_G)$ & $1/R$ (TeV) \\
\hline \hline
      &  $\pi$ & $5.43$ \\
$0.1$ &  $10$ & $3.81$ \\
      &  $50$ & $2.33$  \\
      &  $100$ & $1.88$  \\
      &  $200$ & $1.52$  \\
\hline
      &  $\pi$ & $5.18$ \\
$1.0$ &  $10$ & $3.64$ \\
      &  $50$ & $2.22$  \\
      &  $100$ & $1.80$  \\
      &  $200$ & $1.45$  \\
\hline
      &  $\pi$ & $5.07$ \\
$3.0$ &  $10$ & $3.56$ \\
      &  $50$ & $2.17$  \\
      &  $100$ & $1.76$  \\
      &  $200$ & $1.42$ \\
\hline
      &  $\pi$ & $4.95$ \\
$10.0$ &  $10$ & $3.47$ \\
      &  $50$ & $2.12$  \\
      &  $100$ & $1.71$  \\
      &  $200$ & $1.39$ \\
\hline
\end{tabular}\end{center}
\caption{Some typical solutions of unification.}
\label{tab:RG_solutions}
\end{table}

\section{Decays of KK modes}
Next, we study the decays of KK excitations and begin with KK
photons.

A KK photon  can decay into any charged brane fermions and lower
charged KK fields if  kinematics and KK number selection rules allow it.
Firstly, we will discuss the case for it to decay into brane fermions.
The process is governed by  the effective  lagrangian:
$  \sqrt{2} e A^\mu_n \bar{f} \gamma_\mu f$
which leads to the decay rate
\beq
\Gamma^{f\bar{f}}_A= {N_c \over 12 \pi}(\sqrt{2} e q_f)^2 M_n \sqrt{1-4y_f}(1+2y_f)
\eeq
where $y_f= m_f^2/M_n^2$, $N_c$ is the color factor of fermion $f$
and  $q_f$ is the fermion charge in the unit of electron charge.
Compared with $1/R\sim 1.5-5$TeV all   fermion masses can be ignored,
even that of the t-quark.
Summing over the SM charged fermions, the total decay width is
\beqa
\Gamma^{f\bar{f}}_A &=& 3{2\alpha \over 3}M_n\left[(-1)^2+3\left(\frac23\right)^2
+3\left(-\frac13\right)^2\right]\nonr\\
&=& {16 \alpha\over 3}\frac{2n}{R}\,.
\eeqa

Secondly , we check whether a given KK photon can decay into final states with lighter KK modes.
For the n-th KK photon, the KK number conservation allows it to
decay into a pair of KK $W$ bosons such as   $W_m W_{n-m}$ with the integer $m<n$ or into two channels
with  $U,V$ final states, see appendix for details.
But one of the $U,V$ channel is forbidden by kinematics.
At this stage, we do not  need to consider KK masses splitting
due to  quantum corrections  except to note that they are small.
The tree level  masses relation predicts that $M^A_n \sim M^U_m + M^U_{n-m+1}$
for the decay of $A_n\ra U^{\pm 2}_m+U^{\mp 2}_{n-m+1}$.
The phase space of this process is fully saturated and leave no room for it to happen.
Similar consideration applies to any lower KK number final states at tree level.
We conclude that decays of all KK excitations
are dominated by  the final states with brane fermions.
So the decay widths are the standard ones times two, due to $\sqrt{2}$ factor in
the couplings, with the masses substituted by KK masses:
\beqa
\Gamma_W^{(n)} &=& {(\sqrt{2}g_2)^2 M_{Wn}\over 48 \pi}(3 + 3\sum_{ij}|V_{ij}|^2)
\nonr\\
&=& { 2\alpha\over s_W ^2}\frac{2n}{R}\,,\\
\Gamma_Z^{(n)} &=& { M_{Zn}\over 24 \pi}({\sqrt{2}g_2\over c_W})^2
 \sum_f N_c^f( g_{Lf}^2+ g_{Rf}^2)\nonr\\
&=& { 2\alpha  \over s_W^2 c_W^2}\left(1-2s_W^2+\frac83 s_W^4\right)\frac{2n}{R}\,,\\
\Gamma_V^{(n)}&=& \Gamma_U^{(n)} \sim {(\sqrt{2}g_2/\sqrt{2})^2 M_{Un}\over 24 \pi}\times
3\nonr\\
&=&{ \alpha  \over 2 s_W ^2  }\frac{2n-1}{R}\,,
\eeqa
where $s_W(c_W)$ denotes $\sin\theta_W(\cos\theta_W)$ and
\beq
\Gamma_H^{(n)}= { M_{Hn}\over 4 \pi} \sum_{ij} N_c
{|f_{ij}|^2\over \pi R M^*}
\eeq
for  the KK scalars.  Since the $f_{ij}$ are expected
to be small, we expect  them to  be narrow resonances.
In the above expressions all the SM masses are ignored.

\section{Low energy constraints}
\subsection{Muon Decays}
\begin{figure}[tc]
\centering \includegraphics[width=0.9\columnwidth]{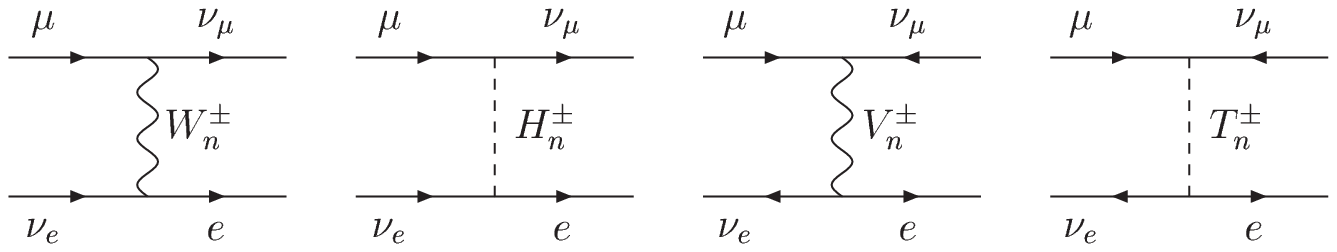}
\caption{The tree-level Feynman diagrams for muon decays}
\label{fig:MuDecay}
\end{figure}

The muon decay is dominated by exchanging $W^\pm,V^\pm$ gauge boson and their
KK excitations, see Fig.\ref{fig:MuDecay}. It also gets  small contributions
from exchanging
physical charged Higgs from the doublet, the triplet and the anti-sextet.
We can ignore these  scalars contribution because they are suppressed by
their Yukawa couplings.

The KK excitations of $W^\pm$ and $V^\pm$ give
extra effective 4-fermion interactions as follows
\beqa
\tri H &\sim& \sum_{n=1} {g_2^2 R^2 }\left[ {1\over (2n)^2}(\bar{\nu}_\mu \gamma^\mu P_L \mu)
(\bar{e}\gamma_\mu P_L \nu_e)\right.\nonr\\
&+&\left. { {\cal U}^*_{ee}{\cal U}_{\mu\mu}\over (2n-1)^2} (\bar{\nu}_\mu \gamma^\mu P_L \mu^\CC)
(\bar{e^\CC}\gamma_\mu P_L \nu_e)\right]\nonr\\
&=& g^2_2{\pi^2R^2\over 24}\left[ (\bar{\nu}_\mu \gamma^\mu P_L \nu_e)
(\bar{e}\gamma_\mu P_L \mu)\right.\nonr\\
&-&\left.  3\,{\cal U}^*_{ee}{\cal U}_{\mu\mu} (\bar{\nu}_e \gamma^\mu P_L \nu_\mu)
(\bar{e}\gamma_\mu P_R  \mu )\right]
\eeqa
where ${\cal U}=U_L^\dag U_R^*$ and $U_L$ is the unitary matrix that
rotates the $SU(2)$ doublet leptons into their mass eigenstates,
as discussed in section II.

The first term adds to the contribution from $W$ boson which is the zero mode
and gives the dominant renormalization of the usual Fermi coupling $G_F^\mu$ \cite{BMar}:
\beq
\label{eq:GF}
G_F^\mu\sim {\sqrt{2}g_2^2 \over 8 M_W^2}\left[1+{\pi^2 M_W^2 R^2\over 12 }\right].
\eeq
This correction is universal for all leptons and quarks.
The contribution of $V$ gives non-zero $g^S_{RR}$ coefficient
\[
 g^S_{RR}= {(\pi M_W R)^2 \over 2}  {\cal U}^*_{ee}{\cal U}_{\mu\mu}
\]
in the notations of \cite{PDG}. It has an upper bound
$| g^S_{RR}|<{(\pi M_W R)^2 \over 2}<0.018$.
This in turn will modify the Michel parameters $\xi$ and $\rho$ to
\beq
\xi = 1- {|g^S_{RR}|^2 \over 4}, \, \rho= \frac34\left(1+{|g^S_{RR}|^2 \over
4}\right).
\eeq
But the deviations from the SM are $-8.1\times 10^{-5}<(\xi-1)<0$ and
$0<(\rho-0.75)< 6.1\times 10^{-5}$. They are beyond the reach of currently
available experiments.
Also the branch ratio of lepton number violation $\mu^+\ra e^+ \bar{\nu_e} \nu_{\mu}$
decay can be estimated to be $\sim | g^S_{RR}|^2 <10^{-4}$. This is insufficient to
account for the LSND neutrino anomaly \cite{LSND}.

\subsection{Muonium-antimuonium conversion and $\mu\ra 3e$}

\begin{figure}[tc]
\centering \includegraphics[width=0.9\columnwidth]{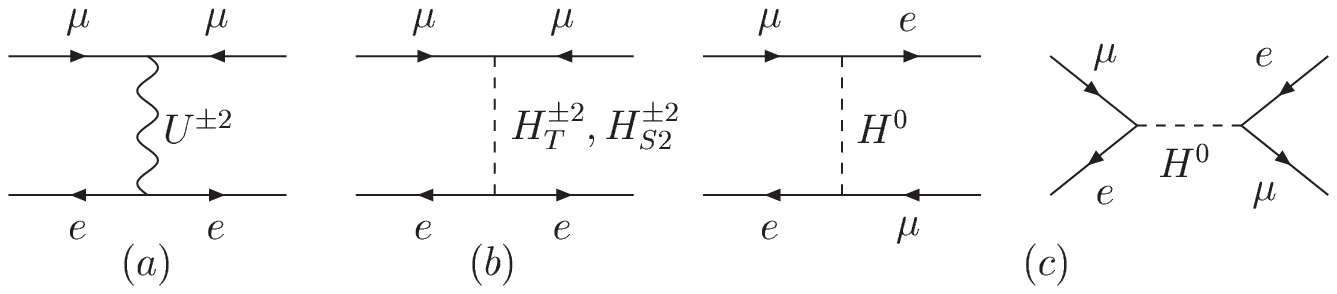}
\caption{ Tree-level diagrams for Muonium-antimuonium conversion.}
\label{fig:MAMC}
\end{figure}

The transition of muonium($M\equiv \mu^+e^-$) into antimuonium ($\overline{M}\equiv \mu^- e^+$)
can  be induced by the exchange of  $U^{\pm2}$ gauge
boson or the scalar KK modes ${H_T}^{\pm2}, H_{S2}^{\pm2}$ which belong to the \sext
( see Fig.\ref{fig:MAMC}(a,b)).
It also can be induced by lepton number conserving but
flavor changing neutral scalars or pseudoscalars as in Fig.\ref{fig:MAMC}(c).

The $M-\overline{M}$ transition amplitude is written in terms of  the mixing
\beq
{\delta\over 2} \equiv \left\langle \overline{M}|H_{M\overline{M}}|M\right\rangle.
\eeq
Using this, the transition possibility is calculated to be
$ P_{M\overline{M}} \sim  \delta^2 / 2 \Gamma_\mu^2$
normalized to muon decay rate $\Gamma_\mu=G_F^2 m_\mu^5/192\pi^3$.
The mixing $\delta$ is sensitive to the helicity structure of the
interaction $H_{M\overline{M}}$ and the total spin of the muonium, for details
see \cite{MAMC}.
 Fig.\ref{fig:MAMC}(a),
gives a  $(V\pm A)(V\mp A)$-type of effective Hamiltonian:
\beq
H^U_{M\overline{M}}= {G^U_{M\overline{M}}\over \sqrt{2}}[\bar{\mu}\gamma^\nu(1-\gamma_5)
e][\bar{\mu}\gamma^\nu(1+\gamma_5)e] + H.c.
\eeq
with the effective Fermi constant given by
\beq
{G^U_{M\overline{M}}\over \sqrt{2}} =  \sum_{n=1}
{ (\sqrt{2}g_2)^2 \over 8  M_{Un}^2}  |{\cal U}^*_{ee}{\cal U}_{\mu\mu}|
= {g_2^2 \pi^2 R^2 \over 32} |{\cal U}^*_{ee}{\cal U}_{\mu\mu}|.
\eeq

 Assuming that the external magnetic field is
zero, we have $\delta^U_3= -8 G^U_{M\overline{M}}/\sqrt{2}\pi a^3$ for the
triplet muonium state and $\delta^U_1= + 24 G^U_{M\overline{M}}/\sqrt{2}\pi a^3$
for the singlet muonium where $a=(\alpha m_e)^{-1}$ is the  Bohr radius.

The relevant lepton number violating scalar interaction,  see Fig. \ref{fig:MAMC}(b), can be
parameterized as
\beq
{\cal L}_{\not L} = \sum_{n=1}\left[ f^T_{ij} \bar{l^\CC}_i P_L l_j H^{+2}_{T,n}
+ f^S_{ ij} \bar{l^\CC}_iP_R l_j H^{+2}_{S2,n}\right] +H.c.
\eeq
which leads to $(V\pm A)^2$-type effective Hamiltonian.
With  the help of Fierz transformations, we obtain the effective Hamiltonian
\beqa
H^{\not L}_{M\overline{M}}= \sum_{n=1}
  f^T_{\mu\mu} f^T_{ee}{[\bar{\mu}\gamma^\nu (1-\gamma_5)e ]^2 \over 8 M^2_{T n} }
\nonr\\
+ \sum_{n=1} f^S_{\mu\mu} f^S_{ee}{[\bar{\mu}\gamma^\nu (1+\gamma_5)e ]^2 \over 8 M^2_{S n} }
 +H.c.
\eeqa
where $f^T= (U_L^T f_6 U_L)/\sqrt{2\pi R M^*}$
and $ f^S = (U_R^T f_6 U_R)/\sqrt{2\pi R M^*}$.
 Since $(V+A)^2$ and $(V-A)^2$ give same contribution to the mixing
matrix element, we can simply add them up  and arrive at
\beq
{G^{\not L}_{M\overline{M}}\over \sqrt{2}}\sim {\pi^2 R^2 \over
64}(f^T_{\mu\mu} f^T_{ ee} +f^S_{\mu\mu} f^S_{ee} ).
\eeq
The lepton number violating interaction due to doubly charged scalars
contributes an amount
$\delta^{\not L}= 16 G^{\not L}_{M\overline{M}}/ \sqrt{2}\pi a^3$  to  the $M-\overline{M}$ mixing
for both muonium singlet and triplet states.
Despite the appearance  note  that the relative sign between contributions from
$U$ gauge boson and $H_{S2}, H_{T}$ scalars is not determined  until the details of
Yukawa couplings and mixing matrix ${\cal U}$ are  known.

Next, we turn our attention to the transition due to flavor changing  neutral
scalars, see Fig.\ref{fig:MAMC}(c).  First, let us  parameterize the interaction as
\beq
{\cal L}_{\rm FC} = f^{\mu e}_i \sum_{i,n=0}\kappa_n \bar{\mu} e H^0_{i n} +
 i f^{\mu e}_{P_i} \sum_{i,n=0}\kappa_n \bar{\mu}\gamma_5 e H^0_{P_i n} + H.c.
\eeq
where $H^0_{i n}$ stands for the n-th KK mode of physical neutral scalar-$i$
and $H_P$ is the physical pseudoscalar-$i$.
In this model, we have two physical neutral scalar zero modes and
one physical pseudoscalar zero mode.
 Generally speaking, one linear combination
of  the fifth component of gauge boson and the two scalar KK modes
will be the Goldstone boson that is the  KK gauge boson longitudinal
component. So we are still left with three physical neutral KK
scalars and three KK pseudoscalar for each KK level from $\phi_3,\phi_3'$ and $\phi_6$. The
linear combination depends on the details of how the 5D gauge is
fixed, such details can be ignored for we just need a qualitative analysis here.

The above lagrangian induces a effective Hamiltonian
\beq
H_{SP}=  {G^S_{M\overline{M}}\over \sqrt{2}}(\bar{\mu}e)^2
- {G^P_{M\overline{M}}\over \sqrt{2}}(\bar{\mu}\gamma_5 e)^2 + H.c.
\eeq
with
\beqa
{G^S_{M\overline{M}}\over \sqrt{2} }
&\sim&
\sum_i (f^{\mu e}_{++i})^2\left[{1\over M^2_{Hi}} + {(\pi R)^2\over 12}\right]
\nonr\\
&+& (f^{\mu e}_{+-})^2 {(\pi R)^2\over 4} \nonr\\
{G^P_{M\overline{M}}\over \sqrt{2}}
&\sim &
(f^{\mu e}_{++P})^2\left[{1\over M^2_{P}} + {(\pi R)^2\over 12}
\right]\nonr\\
&+& \sum_{i}(f^{\mu e}_{+-P_i})^2 {(\pi R)^2\over 4}\nonr
\eeqa
where $M_{Hi}$ and $M_P$ are  the  zero mode masses for $(++)$-parity
scalars and pseudoscalar.

The resulting mixings are
\beq
\delta^{SP}_1= (4 G^S_{M\overline{M}}-8 G^P_{M\overline{M}})/\sqrt{2}\pi a^3
\eeq
for singlet state and
\beq
\delta^{SP}_3= -4 G^S_{M\overline{M}}/\sqrt{2}\pi a^3
\eeq
for the triplet state.

Clearly, more information on the new physics can be obtained  if experiments can be done
with  separated muonium singlet and triplet states.
To make things simple, we will just assume the muonium is prepared in
a statistical mixture; namely $25\%$ is in singlet state and $75\%$ is in
triplet state. Then we derive the  transition probability to be:
\beqa
P(\overline{M})\sim 64^2 \left({6\pi^2 \alpha^3\over G_F m_\mu^2}\right)^2
\left({m_e\over m_\mu}\right)^6 \left({\overline{G}_{M\overline{M}} \over G_F
}\right)^2\nonr\\
= 1.75\times  10^{-6}\left({\overline{G}_{M\overline{M}} \over
G_F}\right)^2
\eeqa
where the effective 4 fermion coupling constant is
\beqa
\overline{G}^2_{M\overline{M}}= \frac14\left[-2 G^U_{M\overline{M}}+ 4 G_{M\overline{M}}^{\not
L}+ G_{M\overline{M}}^S - 2G_{M\overline{M}}^P\right]^2\nonr\\
+\frac34\left[ 6 G_{M\overline{M}}^U + 4 G_{M\overline{M}}^{\not L}-G_{M\overline{M}}^S\right]^2
\,.
\eeqa
The present experimental bound $P_{M\overline{M}}<8.3\times 10^{-11}$
\cite{Willmann:1998gd} requires that $\overline{G}_{M\overline{M}}<6.9\times 10^{-3} G_F$.
Obviously, the actual number of $\overline{G}_{M\overline{M}}$ strongly depend on the pattern of Yukawa
couplings.
For brevity, we will just discuss constraints from two simplified Yukawa
patterns: (1) The diagonal $f_6$ case as discussed in \cite{SU3:triumf} and (2)
the democratic patterns which will be discussed below.

%%%%%%%%%%%%%%%%%%%%%%%%%%%%%%%%%%%%%%%%%%%%%%%%%%%%%%%%%%%%
First, we will consider the diagonal Yukawa pattern with
 $f^6_{ij}\propto  \delta_{ij} m_i/M_W$ and $f_3 \sim f'_3 \ll f^6$.
Since the charged lepton masses are mainly controlled by the
Yukawa couplings of the \sext  all the Yukawa couplings are
expected to be suppressed by $m_l/M_W$ and the transition is mainly due
to $U$ gauge boson.
In this case, the effective $\overline{G}_{M\overline{M}}$ can be simplified to:
\beq
{\overline{G}_{M\overline{M}}\over G_F}\sim \sqrt{28} {G^U_{M\overline{M}}\over G_F}
= -20.93\times 10^{-3}
\left( {2\mbox{TeV}\over 1/R}\right)^2 \times |{\cal U}^*_{ee}{\cal
U}_{\mu\mu}|\,.
\eeq
Here  the lepton mass eigenstates and gauge eigenstates almost coincide,
the mixing is nearly diagonal ${\cal U}^*_{ee}\sim{\cal U}_{\mu\mu}\sim 1$.
To stay under the experimental bound one requires $1/R>3.48$ TeV.
From the discussion of unification we see that such high
compactification scale is not impossible when the factor $\pi R M_G$ is
less than $10$. Note that the unification scale $M_G$ needs not be the fundamental
scale $M^*$. Therefore, even in this case one can still have large volume dilution
factor required by the strong coupling assumption.

But this argument does not apply to the flavor changing channel
mediated by neutral scalars whose Yukawa couplings are  not proportional to
lepton masses. For example, in this model $\phi'_3$ has nothing to do
with charged lepton masses and the flavor changing Yukawa coupling is roughly
of the amount $f'_3|f^3/f^6|$ which is not a severe suppression factor.
Also,  there are two physical neutral scalars
and one physical pseudoscalar which have zero modes with masses around a few hundred GeV.
This is to be compared to  the masses of $U^{\pm2}$ which is around few TeV.
 If these scalars have approximately two orders of magnitude enhanced couplings
they can be as important as the $U$'s.
Moreover, the resulting constraints can be relaxed or tighten depends on the relative sign
between the contribution of vector bileptons and the flavor changing neutral scalars.
Hence, no firm conclusion with regard to these scalars.
We shall proceed by assuming they do not make important contributions.

%%%%%%%%%%%%%%%%%%%%%%%%%%%%%%%%%%%%%%%%%%%%%%%%%%%%%%%%%%%

Next, we study the following so called democratic Yukawa structure whose leading order
is:
\beq
f_6\sim 0.1 \left( \begin{array}{ccc} 1&1&1 \\ 1&1&1 \\ 1&1&1
 \end{array}\right),\hs
f_3\sim 0.01 \left( \begin{array}{ccc} 0&1&1 \\ -1&0&1 \\ -1&-1&0
 \end{array}\right)\,.
\eeq
Our numerical search indicates that it is easy to get realistic solutions which yield observed
charged lepton mass hierarchy and give $|{\cal U}^*_{ee}{\cal U}_{\mu\mu}|<0.1$.
In that case, basically  the $M-\bar{M}$ conversion posts no constraint on $1/R$.
One might wonder if this Yukawa pattern will change neutrino mass
pattern. Correspondingly, we found a simple pattern of $y'_3$ with
$f^{'3}_{12}:f^{'3}_{13}:f^{'3}_{23} \sim 1 : m_\mu/m_\tau : m_e /m_\tau$
which can gives desired bi-large mixing neutrino mass matrix of inverted
hierarchy type.

%%%%%%%%%%%%%%%%%%%%%%%%%%%%%%%%%%%%%%%%%%%%%%%%%%%%%%%%%%%

Nevertheless, we expect the constraint can be loosen once
the contributions from the Higgs sectors are included.
The Yukawa coupling pattern can be arranged such that either the coupling of
$U$ is tiny or the scalars come in and play a significant role to balance
the contribution form $U^{\pm2}$ gauge boson.
To find such a pattern is a  nontrivial task which we shall leave it
to  future investigations and assume $1/R \sim 1.5-3.5$ TeV is viable for the rest
part of the paper.

We note in  passing that the supersymmetrical version of this model will push
the unification scale  $1/R$  to $\sim 6$ TeV\cite{orbifold_SU3}. Which will significantly
suppress this process but  make it  less interesting for collider search.

\begin{figure}[tc]
\centering \includegraphics[width=0.9\columnwidth]{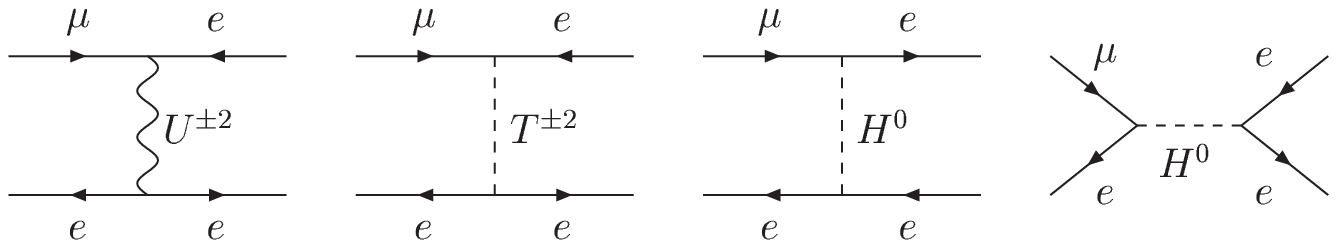}
\caption{Tree-level diagrams for $\mu\ra 3e$.}
\label{fig:Mu_3e}
\end{figure}
\begin{widetext}
The diagrams of  Fig. \ref{fig:Mu_3e} will  lead to $\mu\ra 3e$ transition.
The amplitude is
\beqa
i{\cal M}&=&
 (ig)^2\sum_{n=1}{-2i \over u- M_{Un}^2}
\left[\bar{e}\gamma^\mu{P_L-P_R \over 2}{\cal U}_{ee}e^\CC\right]
\left[\bar{e^\CC}\gamma_\mu { P_L {\cal U}^*_{\mu e} -P_R {\cal U}^*_{e\mu}\over
2}\mu\right]\nonr\\
&+&
 (i\sqrt{2} (f_T)_{e\mu})(i\sqrt{2} (f_T^\dag)_{ee}) \sum_{n=1}{ 2i \over u- M_{Tn}^2}
\left[\bar{e^\CC} P_L \mu \right]\left[\bar{e}P_R e^\CC \right] +
\ldots
\eeqa
where the dots represent the possible flavor changing neutral current
( FCNC ) scalar couplings.
After Fierz rearrangement, we have the effective Lagrangian for the  decay:
\beqa
{\cal L}&=& \sum_{n=1} {g_2^2 {\cal U}_{ee} \over M_{Un}^2}
\left\{
(\bar{e}\gamma^\mu P_L \mu)(\bar{e}\gamma_\mu P_R e){\cal U}^*_{\mu e}
+(\bar{e}\gamma^\mu P_R \mu)(\bar{e}\gamma_\mu P_L e){\cal U}^*_{e\mu}\right\}\nonr\\
 &&+\sum_{n=1} {2 (f_T)_{e\mu}(f_T^*)_{ee} \over M_{Tn}^2}
 [(\bar{\mu}\gamma^\mu P_L e)(\bar{e}\gamma_\mu P_L e)] + H.c.  + \ldots\nonr\\
&=& {(g_2 \pi R)^2 \over 8 }\left\{
(\bar{e}\gamma^\mu P_L \mu)(\bar{e}\gamma_\mu P_R e){\cal U}_{ee}{\cal U}^*_{\mu e}
+(\bar{e}\gamma^\mu P_R \mu)(\bar{e}\gamma_\mu P_L e){\cal U}_{ee}{\cal
U}^*_{e\mu}\right.\nonr\\
&&\left.+2 (f_T)_{e\mu}(f_T^*)_{ee}(\bar{\mu}\gamma^\mu P_L e)(\bar{e}\gamma_\mu P_L e)
\right\}+ H.c.  + \ldots
\eeqa
The branching ratio is given by
\[
Br(\mu\ra 3e) =  {(g_2\pi R)^4\over 512 G_F^2}
\left\{ (|{\cal U}_{\mu e}|^2+ |{\cal U}_{e\mu}|^2) |{\cal U}_{ee}|^2
+ 8 |(f_T)_{e\mu}(f_T^*)_{ee}|^2 +\ldots \right\}\,.
\]
 Note that the scalar contribution here  is positive. For
simplicity, we will only keep the contribution of $U^{\pm2}$ for
estimation:
\beq
Br(\mu\ra 3e)  \sim  1.56\times 10^{-5}\left( 2\mbox{TeV}\over 1/R\right)^4
\left( |{\cal U}_{\mu e}|^2+ |{\cal U}_{e\mu}|^2\right) |{\cal
U}_{ee}|^2\,.
\eeq
The experimental limit $Br(\mu\ra 3e)<1.0\times
10^{-12}$ leads to the requirement that the mixing combination
$( |{\cal U}_{\mu e}|^2+ |{\cal U}_{e\mu}|^2) |{\cal U}_{ee}|^2$  to be very small.
In the diagonal $f_6$ case, ${\cal U}_{ee}\sim 1$ but the mixing ${\cal U}_{\mu e}$ and
${\cal U}_{e\mu }$ are suppressed  by a factor of $\sim (f_3/f_6)$ of small $\phi_3$ Yukawa
coupling. Since theory consistence set a upper bound of $1/R\sim 5$ TeV,
the FCNC couplings must satisfy
$|{\cal U}_{\mu e}|, |{\cal U}_{\mu e}|<1.1\times 10^{-3}$.
On the other hand,  in the democrat Yukawa case the mixing can be made very
small so as  to evade this constraint.
This demonstrates that  in future model building these constraints has to be taken into
account and  Yukawa
couplings are  not totally arbitrary.

Similarly, we have the following  rare decay branching ratios for $\tau$:
\beqa
Br(\tau\ra 3 \mu)&=& 1.56\times 10^{-5}\left( 2\mbox{TeV}\over 1/R\right)^4
 \left(|{\cal U}_{\tau\mu }|^2+ |{\cal U}_{\mu\tau}|^2\right) |{\cal U}_{\mu\mu}|^2\,,\\
Br(\tau\ra 3 e)&=& 1.56\times 10^{-5}\left( 2\mbox{TeV}\over 1/R\right)^4
 \left(|{\cal U}_{\tau e}|^2+ |{\cal U}_{e\tau}|^2\right) |{\cal U}_{ee}|^2\,,\\
Br(\tau\ra  \bar{\mu} ee)&=& 1.56\times 10^{-5}\left( 2\mbox{TeV}\over 1/R\right)^4
 \left(|{\cal U}_{\tau \mu}|^2+ |{\cal U}_{\mu\tau}|^2\right)
|{\cal U}_{e e}|^2\,,\\
Br(\tau\ra  \mu\mu \bar{e})&=& 1.56\times 10^{-5}\left( 2\mbox{TeV}\over 1/R\right)^4
 \left(|{\cal U}_{\tau e}|^2+ |{\cal U}_{e\tau}|^2\right)
|{\cal U}_{\mu \mu}|^2\,,\\
Br(\tau\ra  \mu e\bar{e})&=& 1.95\times 10^{-6}\left( 2\mbox{TeV}\over 1/R\right)^4
 \left(|{\cal U}_{\tau e}|^2+ |{\cal U}_{e\tau}|^2\right)
\left(|{\cal U}_{e\mu}|^2+|{\cal U}_{\mu e}|^2\right)\,,\\
Br(\tau\ra   e\mu \bar{\mu})&=& 1.95\times 10^{-6}\left( 2\mbox{TeV}\over 1/R\right)^4
 \left(|{\cal U}_{\tau \mu}|^2+ |{\cal U}_{\mu\tau}|^2\right)
\left(|{\cal U}_{e\mu}|^2+|{\cal U}_{\mu e}|^2\right)\,.
\eeqa
\end{widetext}
But the current limit $\sim 3\times 10^{-6}$ does not impose strong
constraint on the mixing.
 Because the scalar sector gives positive contribution to
these rare decays,
from the unitarity of ${\cal U}$ the model predicts an interesting  lower bond for a given $1/R$
\beq
Br(\tau\ra 3 e) > 3.12\times 10^{-5}\left( 2\mbox{TeV}\over 1/R\right)^4
  |{\cal U}_{ee}|^2 \left(1-|{\cal U}_{ee }|^2 \right)\, .
\eeq
If one wants to keep compactification scale $1/R$ low, say $\sim
1.5$ TeV, it is  required that $|{\cal U}_{ee}|$ to be either close
to zero or one. Furthermore, if we take the upper bound of $1/R < 5$ TeV derived from
unification seriously we obtain
\beq
Br(\tau\ra 3 e) > 8.0\times 10^{-7}
  |{\cal U}_{ee}|^2 \left(1-|{\cal U}_{ee }|^2 \right)\, .
\eeq
On the other hand, if we assume that the bilepton exchange is the dominate FCNC source,
another interesting upper bond can be derived:
\beq
Br(\tau\ra 3 e) < 7.8\times 10^{-6}\left( 2\mbox{TeV}\over 1/R\right)^4.
\eeq
An observation of this decay will shed light on the Yukawa structure.
%%%%%%%%%%%%%%%%%%%%%%%%%%%%%%%%%%%%%%%%%%%%%%%%%%%%%%%%%%%%%%%%%%%%%%%%%%%%%%
\section{Collider Signatures }
In collider experiments the bilepton signals can be directly
probed. For simplicity, in the following discussion we will not
consider the contribution from scalar bileptons due to the Yukawa suppression.

\subsection{$l^+ l^-$ scattering}

\begin{figure}[tc]
\centering \includegraphics[width=0.9\columnwidth]{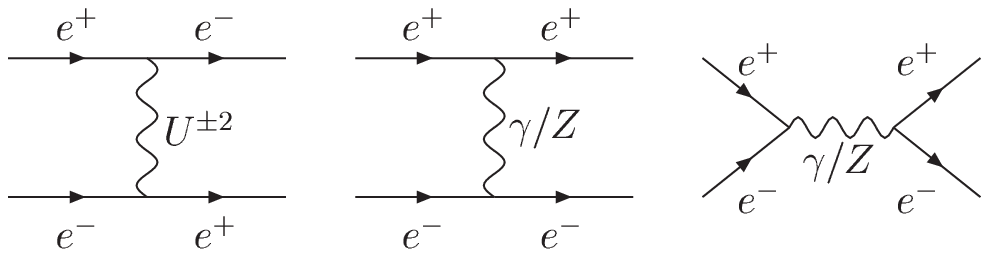}
\caption{Feynman diagrams for $e^+ e^-$ scattering.}
\label{fig:FD_Bhabha}
\end{figure}
\begin{widetext}
First we consider the high energy Bhabha scattering, $e^-(p_1)e^+(p_2)\ra e^-(p_3)
e^+(p_4)$, see Fig.\ref{fig:FD_Bhabha}.
The  amplitude from $u$-channel $U^{\pm2}$ KK tower exchange is given by:
\beq
i {\cal M}_{U} =\sum_{n=1}
{ -i(i g_2)^2 (2!)^2|{\cal U}_{ee}|^2\over u- M_n^2 +i M_n \Gamma_n}
\left( \bar{u}(p_3) \gamma_\mu {P_L-P_R \over 2 }C \bar{v}^T(p_2)\right)
\left( -v^T(p_4)C^{-1} \gamma^\mu {P_L-P_R\over 2}
u(p_1)\right)\,.
\eeq
By Fierz transformation, it can be rearranged to
\beqa
{\cal M}_{U} =\left. \sum_{n=1}
{ - g_2^2 |{\cal U}_{ee}|^2\over u- M_n^2 +i M_n \Gamma_n}
\right[ \bar{u}(p_3) \gamma_\mu P_L  u(p_1) \bar{v}(p_2) \gamma^\mu P_R
v(p_4)
 + \bar{u}(p_3) \gamma_\mu P_R  u(p_1) \bar{v}(p_2) \gamma^\mu P_L
 v(p_4)\nonr\\
\left.- \bar{u}(p_3) \gamma_\mu P_L  v(p_4) \bar{v}(p_2) \gamma^\mu P_R
u(p_1)
- \bar{u}(p_3) \gamma_\mu P_R  v(p_4) \bar{v}(p_2) \gamma^\mu P_L u(p_1)
\stackrel{\mbox{$\,$ }}{\mbox{}} \right]\,.
\eeqa
\end{widetext}
The minus signs in front of the last two terms  are due to Fermi statistics.
%As usual, we denote $s=(p_1+p_2)^2$, $t=(p_1-p_3)^2= -\frac{s}{2}(1-\cos\theta)$
%and $u=(p_1-p_4)^2=-\frac{s}{2}(1+\cos\theta)$.
We use the Mandelstam variables in the following and
combine all the contribution from $U, \gamma, Z$ and their KK excitations.
The total transition amplitude can be written as
\beqa
&&{\cal M}(e^+ e^-\ra e^+ e^-)=\nonr\\
&& g_2^2 G^B_{\lambda,\lambda'} \bar{u}(p_3)\gamma^\mu P_\lambda
v(p_4)\bar{v}(p_2)\gamma_\mu P_{\lambda'} u(p_1)\nonr\\
&-& g_2^2 F^B_{\lambda,\lambda'} \bar{v}(p_2)\gamma^\mu P_\lambda
v(p_4)\bar{u}(p_3)\gamma_\mu P_{\lambda'} u(p_1)
\eeqa
where $\lambda, \lambda' = L, R$.
%Again, extra minus sign need to be taken into consideration.
The coefficients are the sums of all KK excitation as well as SM gauge bosons and  they read
\beqa
G^B_{\lambda,\lambda'}&=& s_W^2 \sum_{n=0}{\kappa_n^2\over s- M_{\gamma n}^2+ i M_{\gamma n}\Gamma_{\gamma n}}
\nonr\\
&+&{g_\lambda g_{\lambda'} \over c_W^2} \sum_{n=0} {\kappa_n^2 \over s -M_{Z n}^2+i M_{Z n}\Gamma_{Zn}}
 \nonr\\
&-& (1-\delta_{\lambda,\lambda'})\sum_{n=1} {  g^B_{\lambda,\lambda'}\over u -M_{U n}^2+i M_{U n}\Gamma_{U n}}
\eeqa
where $ g^B_{\lambda,\lambda'}=|{\cal U}_{ee}|^2$, $g_L = s_W^2-1/2$ and $g_R= s_W^2$.
Similarly,
\beqa
F^B_{\lambda,\lambda'}&=& s_W^2 \sum_{n=0}{\kappa_n^2\over t- M_{\gamma n}^2+ i M_{\gamma n}\Gamma_{\gamma
n}}\nonr\\
&+& {g_\lambda g_{\lambda'} \over c_W^2} \sum_{n=0} {\kappa_n^2 \over t -M_{Z n}^2+i M_{Z n}\Gamma_{Z
  n}}\nonr\\
 &-& (1-\delta_{\lambda,\lambda'})\sum_{n=1} { f^B_{\lambda,\lambda'} \over u -M_{U n}^2+i M_{U n}\Gamma_{U
 n}}
\eeqa
with  $ f^B_{\lambda,\lambda'}=|{\cal U}_{ee}|^2$.
%%%%%%%%%%%%%%%%%%%%%%%%%%%%%%%%%%%%%%%%%%%%%%%%%%%%%%%%%%%%%%%%%

\begin{figure}[tc]
\centering \includegraphics[width=0.9\columnwidth]{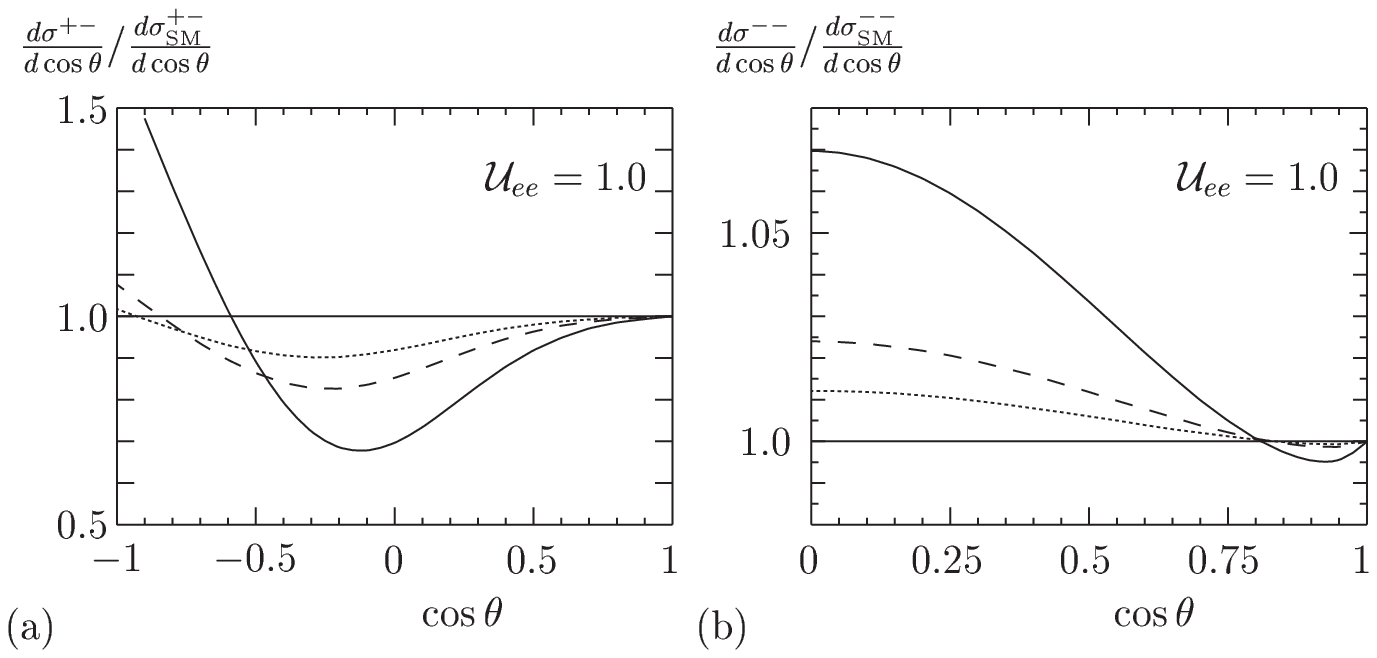}
\caption{The ratio of differential cross section to that of SM at $\sqrt{s}=500$ GeV%
for(a) the Bhabha and (b)M\o ller scattering. The mixing ${\cal U}_{ee}$ is set to 1.%
Solid:$1/R= 1.5$ TeV, Dash:$1/R= 2.5$ TeV; Dotted: $1/R= 3.5$
TeV.}
\label{fig:ADD_10}
\end{figure}

\begin{figure}[tc]
\centering \includegraphics[width=0.9\columnwidth]{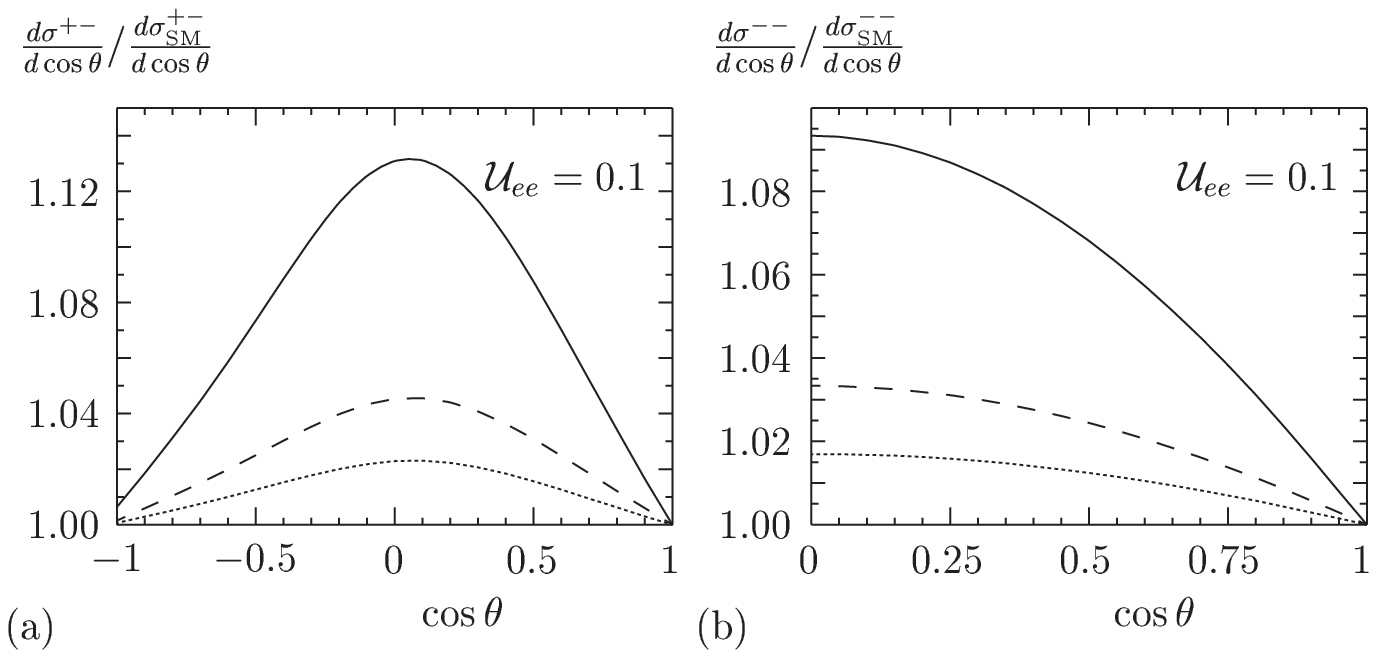}
\caption{ Same as previous fig.\ref{fig:ADD_10}
but ${\cal U}_{ee}=0.1$.}
\label{fig:ADD_01}
\end{figure}

\begin{figure}[tc]
\centering \includegraphics[width=0.9\columnwidth]{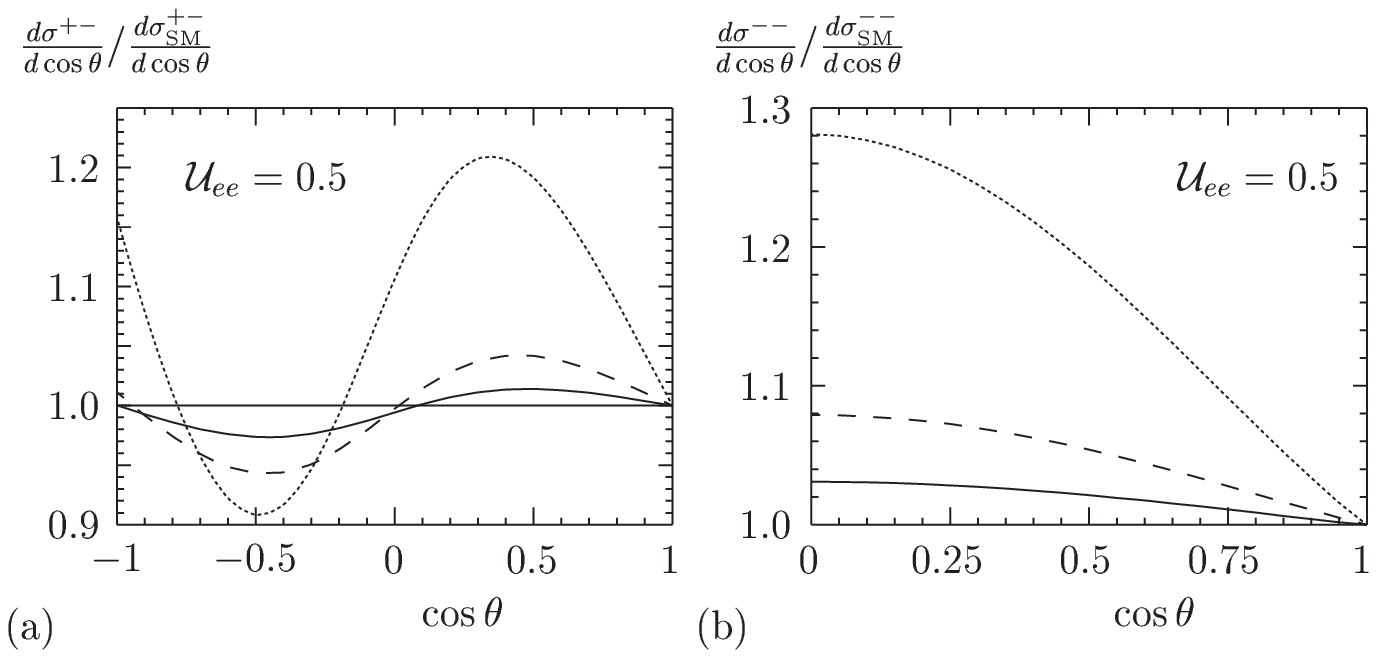}
\caption{Same as figure \ref{fig:ADD_10} with $1/R=2.5$TeV and ${\cal U}_{ee}=0.5$.
The center of mass energy of collider $\sqrt{s}= 500$ GeV(Solid), $800$ GeV(Dash)
and $1.5$TeV(Dotted).}
\label{fig:ADD_05}
\end{figure}

%%%%%%%%%%%%%%%%%%%%%%%%%%%%%%%%%%%%%%%%%%%%%%%%%%%%%%%%%%%%%%%%%%%%%%%%%%%%%%%
The differential cross section can be calculated straightforwardly
\beqa
{d \sigma^{+-} \over d \cos\theta} &=& {\pi \alpha^2 s \over 2s_W^4}\left[
|F^B_{RL}|^2+ |F^B_{LR}|^2 +(|G^B_{RL}|^2+|G^B_{LR}|^2){t^2\over
s^2}\right.\nonr\\
&+&\left.{u^2\over s^2}(|G^B_{LL}+F^B_{LL}|^2+|G^B_{RR}+F^B_{RR}|^2) \right]
\eeqa
with $F^B_{LR}=F^B_{RL}$ and $G^B_{LR}=G^B_{RL}$.
The deviation from the SM prediction are displayed in
the panel-(a) of Figs.\ref{fig:ADD_10}, \ref{fig:ADD_01}, \ref{fig:ADD_05}
for some typical
parameter sets.

Of course, in $e^+ e^-$ experiments, a signal of flavor changing scattering
$e^+e^-\ra l^+_i l^-_j, i\neq j$ is clearly beyond SM. In this model, it
could be mediated by flavor changing coupling of $U^{\pm2}$ gauge
bosons. Also, there is possible small contribution from
$t-$channel FCNC Higgs and scalar bilepton diagrams.
 From our discussion above, the cross section is already at
hand. We just substitute it with the following nonzero variables
\beqa
F^B_{\lambda,\lambda'} =  - \sum_{n=1}
 { (1-\delta_{\lambda,\lambda'})f^B_{\lambda,\lambda'} \over u -M_{U n}^2+i M_{U n}\Gamma_{U
 n}}\nonr\\
G^B_{\lambda,\lambda'} =  - \sum_{n=1}
 {(1-\delta_{\lambda,\lambda'}) g^B_{\lambda,\lambda'} \over u -M_{U n}^2+i M_{U n}\Gamma_{U
 n}} \nonr
\eeqa
in different cases:
%otherwise zero,
\beqa
i,j\neq e  &:& f^B_{RL}={{\cal U}^*_{ei}{\cal U}_{je}\over 4},\,
 f^B_{LR}={ {\cal U}^*_{ie}{\cal U}_{ej}\over 4},\nonr\\
 && g^B_{RL}={{\cal U}^*_{ei}{\cal U}_{ej}\over 4},\,
 g^B_{LR}={{\cal U}^*_{ie}{\cal U}_{je}\over 4}\, ,\\
j=e &:& f^B_{RL}=g^B_{RL} =   { {\cal U}_{ee}{\cal U}^*_{ei} \over
2},\nonr\\
&&f^B_{LR}=g^B_{LR} =   { {\cal U}_{ee}{\cal U}^*_{ie} \over 2}\, ,\\
i=e &:&f^B_{RL}=g^B_{LR} =   { {\cal U}^*_{ee}{\cal U}_{je} \over
2},\nonr\\
&&f^B_{LR}=g^B_{RL} =   { {\cal U}^*_{ee}{\cal U}_{ej} \over 2}\,
.
\eeqa
Then the different flavor violating scattering channels can be
used to fix  off-diagonal entities of mixing matrix $\cal U$.

\subsection{$l^-l^-$ scattering}

\begin{figure}[tc]
\centering \includegraphics[width=0.8\columnwidth]{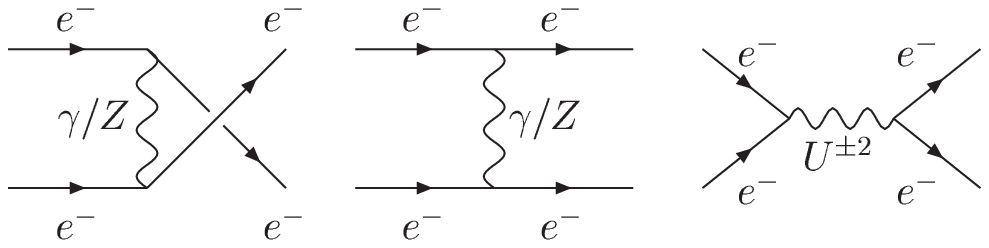}
\caption{Diagrams for $e^- e^-$ scattering.}
\label{fig:FD_Moller}
\end{figure}

Similarly, the amplitude for M\o ller scattering, see fig.\ref{fig:FD_Moller}, is expressed as
\beqa
&&{\cal M}(e^- e^-\ra e^- e^-) =\nonr\\
&& g_2^2
G^M_{\lambda,\lambda'}\bar{u}(p_4)\gamma^\mu P_\lambda u(p_2)
\bar{u}(p_3)\gamma_\mu  P_{\lambda'} u(p_1)\nonr\\
&-& g_2^2
F^M_{\lambda,\lambda'}\bar{u}(p_3)\gamma^\mu P_\lambda u(p_2)
\bar{u}(p_4)\gamma_\mu  P_{\lambda'} u(p_1)
\eeqa
where
\beqa
F^M_{\lambda,\lambda'}&=& s_W^2 \sum_{n=0}{\kappa_n^2\over u- M_{\gamma n}^2+ i M_{\gamma n}\Gamma_{\gamma n}}
\nonr\\
&+& {g_\lambda g_{\lambda'} \over c_W^2} \sum_{n=0} {\kappa_n^2 \over u -M_{Z n}^2+i M_{Z n}\Gamma_{Z
  n}}
\nonr\\
&-& (1-\delta_{\lambda,\lambda'})\sum_{n=1} {f^M_{\lambda,\lambda'} \over s -M_{U n}^2+i M_{U n}\Gamma_{U n}}
\eeqa
and
\beqa
G^M_{\lambda,\lambda'}&=& s_W^2 \sum_{n=0}{\kappa_n^2\over t- M_{\gamma n}^2+ i M_{\gamma n}\Gamma_{\gamma
n}}\nonr\\
 &+& {g_\lambda g_{\lambda'} \over c_W^2} \sum_{n=0} {\kappa_n^2 \over t -M_{Z n}^2+i M_{Z n}\Gamma_{Z
  n}}\nonr\\
 &-& (1-\delta_{\lambda,\lambda'})\sum_{n=1} {g^M_{\lambda,\lambda'} \over s -M_{U n}^2+i M_{U n}\Gamma_{U n}}
\eeqa
with $f^M_{\lambda,\lambda'}=g^M_{\lambda,\lambda'}=|{\cal U}_{ee}|^2$.
The differential cross section is
\beqa
{d \sigma^{--} \over d \cos\theta} = {\pi \alpha^2 s\over 4 s_W^4}\left[
 |G^M_{LL}+F^M_{LL}|^2 + |G^M_{RR}+F^M_{RR}|^2\right.\nonr\\
\left.+ {u^2 \over s^2}(|G^M_{LR}|^2+|G^M_{RL}|^2)
+ {t^2 \over s^2}(|F^M_{LR}|^2+|F^M_{RL}|^2)\right]\,.
\eeqa
The angular distribution is also different from the SM prediction,
see   panel-(b) of Figs.\ref{fig:ADD_10}, \ref{fig:ADD_01}, \ref{fig:ADD_05}
 illustrated  with some typical parameter sets.
Besides comparing  the angular distribution with the SM
prediction,  the flavor changing production $e^- e^-\ra l'^-l^-,\, (l',l\neq e)$ in a
linear collider will be a very clean signal which can only be due to bilepton exchange.
For example, the $e^- e^- \ra \mu^- \mu^-$ cross section is given by
\beq
{d \sigma \over d \cos\theta}= {g_2^4 \over 32 \pi
s}(1+\cos\theta^2)
\left|\sum_{n=1} {s {\cal U}^*_{ee}{\cal U}_{\mu\mu} \over s-M_{Un}^2 + i M_{Un}  \Gamma_{Un} }\right|^2
\label{eq:e-e-m-m-}
\eeq
where $\cos\theta$ goes from $0$ to $1$ due to the final states
are identical particles. If we take ${\cal U}^*_{ee}{\cal U}_{\mu\mu}=1$,
 the total cross section  is $(2.5, 0.3, 0.1)fb$ at $\sqrt{s}=100$
GeV for  $1/R=(1.5,2.5,3.5)$TeV respectively,
 but  increase to  $(74.1,8.4, 2.1)$fb at $\sqrt{s}=500$ GeV and
$(666.7, 42.2, 9.5)fb$ at $\sqrt{s}=1$ TeV.
The same expression, with the appropriate substitutions for the
mixing factors, applies also to $e^-e^-\ra \tau^-\tau^-$
and $\mu^-\mu^-\ra \tau^-\tau^-$ cross sections.
These  will be  unmistakable existence signals of bilepton $U^{\pm2}$.
For other flavor changing scattering,  $e^- e^- \ra e^- \mu^-, \tau^-
\mu^-, \cdots$ etc, it is easy to get the corresponding expression
from Eq. (\ref{eq:e-e-m-m-}).

\subsection{Gauge bosons pair  production signature in colliders}
\begin{figure}[tc]
\centering \includegraphics[width=0.6\columnwidth]{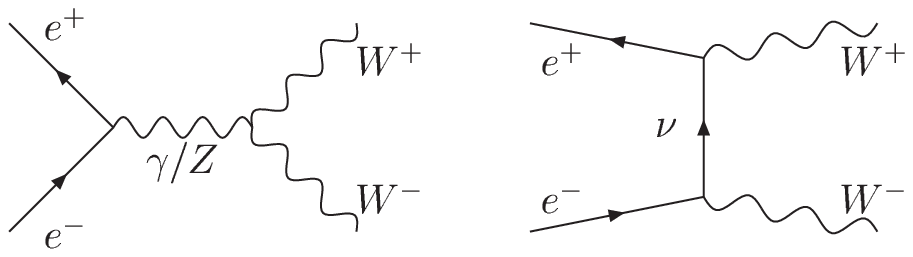}
\caption{Diagrams for $W$ pair production in $e^+ e^-$ collider.}
\label{fig:WPair}
\end{figure}

The SM $W$-pair production is an important test of the SM. The tree level
diagrams are shown in Fig.\ref{fig:WPair} where we do not display  the Higgs
exchange graph.
Here we discuss how will it be affected in the $SU(3)_W$ model.
The first  correction is due to exchanging  KK photons and KK $Z$
bosons. But since the two final state $W^\pm$ are zero modes the
KK number conservation at the triple gauge vertex forces the virtual
neutral gauge boson
to be a zero mode too.
Since the  fermions are 4D brane fields in this model, nothing will be added to
the $t-$channel neutrino diagram. Thus, we conclude there is no tree-level correction
to the SM $W$-pair production process.

Next, can  the $W_1^\pm W^\mp$ be produced in a linear collider?
This can proceed through the $s$ channel photon and $Z$ exchange $e^+e^-\ra \gamma_1/Z_1 \ra W_1^\pm W^\mp$
 and the $t$-channel neutrino exchanging
diagrams. Both are allowed by KK number conservation. But the  $W_1$ mass $\sim 2/R$
requires a linear collider with $\sqrt{s}>3$ TeV.
If the collider is available, we shall also observe
the pair  production of the first KK modes of $U^{\pm2}$ or $V^\pm$ gauge bosons.
For example, $e^+e^-\ra U^{+2}_1 U^{-2}_1$ can be mediated by
the zero modes and the first KK excitation of photon and $Z$  with
distinctive signatures of $U^{\pm2}$ bileptonic decays.

The pair production of bileptons can also be searched in the
hadron collider through virtual photon or $Z$
coming from  $\bar{q}q \ra \gamma^*/Z^*\ra U^{+2}U^{-2}$  or
virtual $W$ from $\bar{q}q'\ra W^* \ra U V$.
For completeness, we also give triple gauge couplings in the
appendix. Due to the high masses involved, a detail phenomenological analysis
involving  hadron machines is premature now.

\section{ $\nu e$ scattering}
\begin{figure}[tc]
\centering \includegraphics[width=0.6\columnwidth]{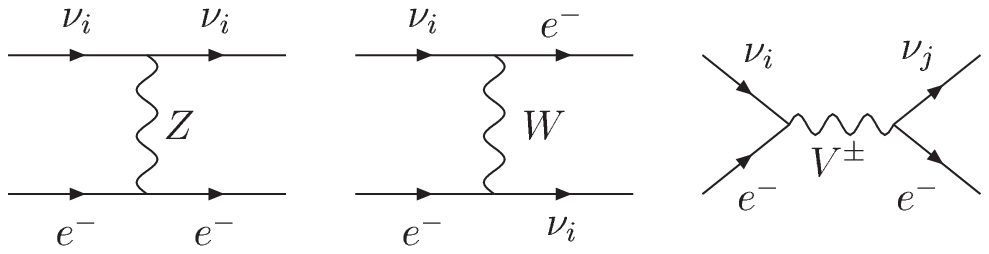}
\caption{The Feynman diagrams for $\nu e$ scattering.}
\label{fig:FDnue}
\end{figure}
The low energy neutrino electron scattering is another place to look
for new physics signals. The $\nu^i e\ra \nu^j e$
 process receives
 corrections form KK modes of $Z, W^\pm$ and $V^\pm$ gauge bosons, see Fig.\ref{fig:FDnue}.
 The scattering amplitude reads
\begin{widetext}
\beqa
i{\cal M}&=&
\sum_{n=0}\left({ig_2 \kappa_n\over c_W}\right)^2 {-i \over (p_1-p_3)^2-M_{Zn}^2}
(\bar{\nu}_i\gamma^\mu g_L^\nu P_L\nu_i)
(\bar{e}\gamma_\mu[g_L P_L +g_R P_R]e)\nonr\\
&+&
\sum_{n=0}\left({ig_2 \kappa_n\over \sqrt{2}}\right)^2 {-i \over (p_1-p_4)^2-M_{Wn}^2}
(\bar{\nu}_i\gamma^\mu P_L \nu_i)(\bar{e}\gamma_\mu P_L e)
+\sum_{n=1}\left({ig_2 \kappa_n\over \sqrt{2}}\right)^2 {-i {\cal U}_{i e}{\cal U}^\dag_{j e} \over (p_1+p_3)^2-M_{Vn}^2}
(\bar{\nu}_j\gamma^\mu P_L e^\CC)
(\bar{e^\CC}\gamma_\mu P_L \nu_i)\,.\nonr
\eeqa
The momenta transfers can be ignored compared to the gauge boson
masses we can rewrite it as
\beqa
 {\cal M}&\simeq&
-{g_2^2 \over 2 c_W^2}\left(\frac{1}{M_Z^2}+{\pi^2R^2\over 12} \right)
(\bar{\nu}_i \gamma^\mu P_L \nu_i)(\bar{e}\gamma_\mu[ g_L P_L+ g_R P_R]
e)\nonr\\
&-&
{g_2^2 \over 2}\left(\frac{1}{M_W^2}+{\pi^2R^2\over 12} \right)
(\bar{\nu}_i \gamma^\mu P_L \nu_i)(\bar{e}\gamma_\mu P_L e)
+{g_2^2 \over 2 } {{\cal U}_{i e}{\cal U}^\dag_{j e}\pi^2R^2\over 4}
(\bar{\nu}_j \gamma^\mu P_L \nu_i)(\bar{e}\gamma_\mu P_R e)\,.
\eeqa
\end{widetext}
We parameterize it in the standard  effective Hamiltonian form
with  the effective Fermi constant  $G_F^\mu$ given by Eq. (\ref{eq:GF}):
\beq
H_{eff}= {G^\mu_F \over \sqrt{2}}
(\bar{\nu}_i\gamma^\mu(1-\gamma_5)\nu_j)(\bar{e}\gamma_\mu
(g^{ij}_V- g^{ij}_A \gamma_5)e)
\eeq
with
\beqa
g_V^{ee} \sim  (2s_W^2-\frac12 )\left(1+\epsilon t^2_W \right)+1-3\epsilon |{\cal
U}_{ee}|^2\,,\\
g_A^{ee} \sim -\frac12\left(1+\epsilon t^2_W \right)+1+3\epsilon |{\cal U}_{ee}|^2\,,\\
g_V^{\mu\mu} \sim  (2s_W^2-\frac12 )\left(1+\epsilon t^2_W\right)
-3\epsilon |{\cal U}_{\mu e}|^2\,,\\
g_A^{\mu\mu} \sim  -\frac12 \left(1+\epsilon t^2_W\right)
+ 3\epsilon |{\cal U}_{\mu e}|^2\,,\\
g_V^{\tau\tau} \sim (2s_W^2-\frac12 )\left(1+\epsilon t^2_W\right)
- 3\epsilon |{\cal U}_{\tau e}|^2\,,\\
g_A^{\tau\tau} \sim  -\frac12\left(1+\epsilon t^2_W\right)
+3\epsilon |{\cal U}_{\tau e}|^2
\eeqa
where $t_W^2=\sin^2\theta_W/\cos^2\theta_W$ and $\epsilon\equiv(\pi R M_W)^2/12$.
In this model, only $V^\pm$ gauge bosons pay tribute to the neutrino flavor
changing scattering and give
\beq
-g_V^{ji}=g_A^{ji} \sim 3
 \epsilon {\cal U}_{i e}{\cal U}^\dag_{j e}\, ,\,
 i\neq j.
\eeq
If seen, it is clearly physics beyond SM.

\begin{figure}[tc]
\centering \includegraphics[width=0.8\columnwidth]{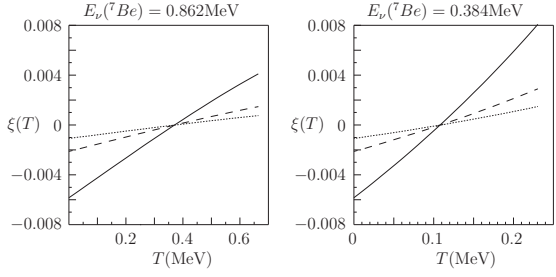}
\caption{The ratio of electron recoil spectrum from $0.862$-MeV  and $0.384$-MeV $^7Be$
neutrino line to the SM tree-level prediction. The mixing ${\cal U}_{ee}$ is set to unity and
$1/R= 1.5, 2.5, 3.5$ TeV for solid, dash and dotted curves
respectively.} \label{plot:nue}
\end{figure}

We are interested in the situation that the incoming neutrino scatters off the
electron  at rest, $\nu^i(E_\nu) +e \ra \nu^j + e(E_e)$,
as been measured  in the current neutrino experiments.
Define the electron recoil energy as $T= E_e - m_e$
then the effective Hamiltonian leads to the following cross section \cite{'tHooft:ht}
in  the regime $q^2\ll M_W^2$
\beqa
{d \sigma \over d T}= {2 G_{\mu }^2 m_e \over \pi}\sum_j
\left\{ (g_L^{ij})^2
+(g_R^{ij})^2\left(1-\frac{T}{E_\nu}\right)^2\right.\nonr\\
\left.-g_L^{ij}g_R^{ij} {m_e T \over E_\nu^2} \right\}
\eeqa
where $g_L^{ij} =(g_V^{ij}+g_A^{ij})/2$ and $g_R^{ij}
=(g_V^{ij}-g_A^{ij})/2$. The sum over final neutrino species is taken as
they are not detected in such experiments.
For incoming anti-neutrino, as in the reactor neutrino experiments,
the cross section for $\bar{\nu}^i e$ scattering is simply
\beqa
{d \sigma \over d T}= {2 G_{\mu }^2 m_e \over \pi}\sum_j
\left\{ (g_R^{ij})^2
+(g_L^{ij})^2\left(1-\frac{T}{E_\nu}\right)^2\right.\nonr\\
\left. -g_L^{ij}
g_R^{ij} {m_e T \over E_\nu^2} \right\}.
\eeqa
Labeling $\phi$ as the angle which recoiled electron is deflected  from the
incident neutrino, it relates to $T$ as
\beq
\cos\phi = {E_\nu +m_e \over E_\nu}\sqrt{{T\over T +2 m_e}}.
\eeq
The electron recoil energy is in the range of $0 \leq T\leq E_\nu^2/(E_\nu
+m_e/2)$.
We denote  $\xi(T)$ as the ratio of the modified differential cross
section to the SM as follows
\beq
{d \sigma \over d T}= {d\sigma_{SM} \over d T}(1+\xi(T)).
\eeq
For $|{\cal U}_{ee}|=1$, the plot of $\xi(T)$ vs $T$ for two $^7Be$ neutrino lines
is  given in  fig.\ref{plot:nue}.
This model gives compatible but opposite corrections to the SM
radiative correction \cite{Bahcall:1995mm} on the electron recoil energy spectrum.
This  could be searched for in future neutrino experiments.
For zero mixing $|{\cal U}_{ee}|=0$, $\xi(T)$ is nearly flat, but not zero due to
KK photon and $Z$. The value of  $\xi(T)$ is reduced  to $-0.0008, -0.0003$
and $-0.0002$ for both neutrino lines at $1/R=1.5, 2.5, 3.5$ TeV respectively.

For neutrino source with continuous energy distribution, the averaged
differential scattering cross section as a function
of $T$ is experimentally interesting.
\beq
\left\langle {d\sigma \over dT}\right\rangle_T =\int^{E_{max}}_{E_{min}} d
E_\nu \lambda(E_\nu) {d\sigma \over dT}
\eeq
where $\lambda(E_\nu)$ is the probability distribution in terms of neutrino
energy. The minimum energy of incident neutrino to give electron
recoil energy $T$ is
\beq
E_{min}={T+\sqrt{T(T+2m_e)} \over 2}.
\eeq
Similarly, we define
\beq
\left\langle {d\sigma \over d T}\right\rangle_T \equiv
\left\langle {d\sigma_{\rm SM} \over d
T}\right\rangle_T(1+\bar{\xi}(T))\,.
\eeq

\begin{figure}[tc]
\centering \includegraphics[width=0.9\columnwidth]{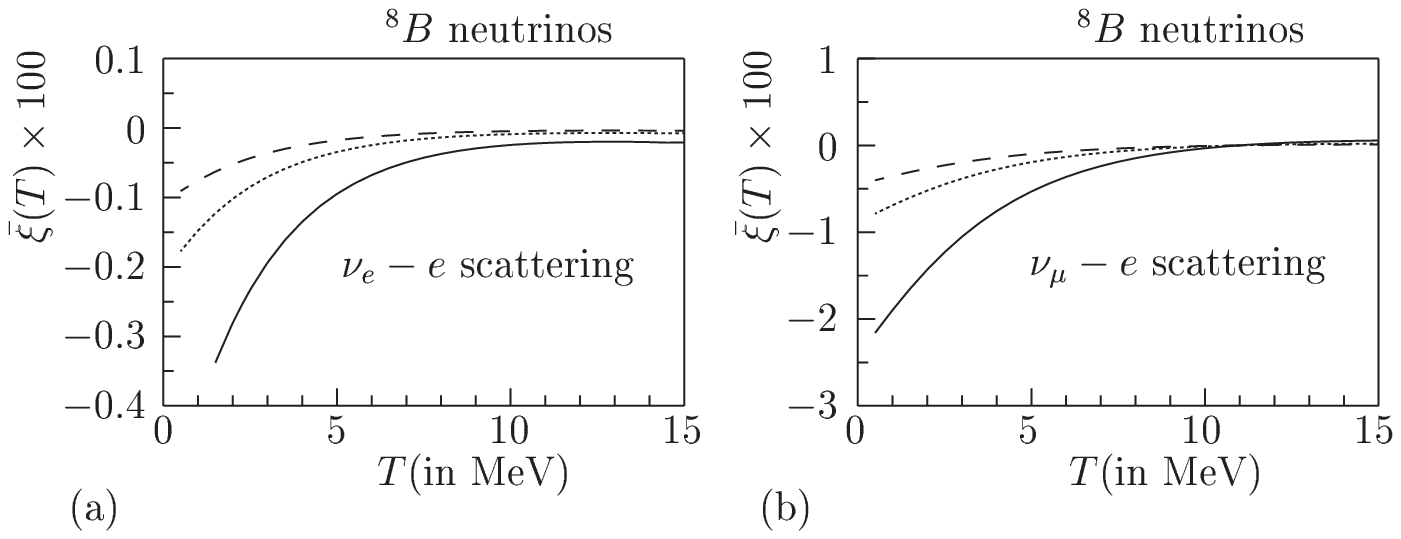}
\caption{The  recoil spectrum distortion for (a) $\nu_e-e$ and (b) $\nu_\mu-e$ scattering
from solar $^8B$ neutrino.
The  mixing $|{\cal U}_{ee, \mu e}|$ is set to unity for (a,b)  and
$1/R= 1.5, 2.5, 3.5$ TeV for solid, dash and dotted curves
respectively.} \label{plot:nuB8}
\end{figure}

For water Cerenkov type experiments, the dominate solar neutrinos
come from the decay process $^8B\ra\, ^7Be^* +e^+ +\nu_e$ and
the small fraction of $hep$ neutrinos from process
$^3He+p\ra\, ^4He +e^+ +\nu_e$ can be ignored.
We give plots on $\bar{\xi}(T)$ for solar $^8 B$ neutrinos in
Fig.\ref{plot:nuB8} with the $\lambda(E_\nu)$ adopted form \cite{BahcallB}.
In the best case scenario, i.e. low $1/R$ and maximal mixing, the deviation
could reach level of a percent at low recoil energies. This is comparable to
the  radiative  correction suppression  at high recoil energies \cite{Bahcall:1995mm}.
When the mixing vanishes, vector bilepton $V^\pm$ has no
contribution, we are left with the effect from KK $W$ and $Z$ which gives
a roughly constant correction $\bar{\xi}(T)\sim \{-0.0005, -0.0002, -0.0001\}
, \{0.0014, 0.0005, 0.0003\}$ for $\nu_{e,\mu}-e$ scattering
with $1/R=\{1.5,2.5,3.5\}$ TeV respectively.
Note that unitarity condition requires $|{\cal U}_{ee}|^2+|{\cal U}_{\mu e}|^2
+|{\cal U}_{\tau e}|^2 = 1$, so the bilepton effects always show up
in some flavor combination.

\section{Conclusion}
We have carefully gone through  the  possible phenomenology of the
5D orbifold $SU(3)_W$ GUT model \cite{SU3:triumf}.
The RGEs and unification has been examined in section III, we found that the
compactification scale, $1/R$, is around $1.5-5$ TeV which makes
phenomenology interesting.

Due to KK number conservation and tree-level mass relations,
the decays of KK excitations are dominated by the final states
constituted by brane fermions, i.e. the SM fermions, only.
At the tree level, this results in a universal prediction of
 $\Gamma^{KK}_n /M^{KK}_n \sim 2 \Gamma^{SM}/M_{SM}$ for any KK state
 and level $n$.

The scalar and vector bileptons come naturally with the $SU(3)_W$ gauge
symmetry, which induces many testable signatures.
The couplings between vector bilepton and leptons are  modulated by
a CKM like unitary mixing matrix , ${\cal U}$, which is  controlled by the
details of Yukawa sector. In principle, the ${\cal U}$ can give CP violation in
the lepton sector which will be  addressed elsewhere.
Lepton flavor changing processes stem from non-diagonal ${\cal U}$.
Among them, we found that the muonium-antimuonium conversion
and $\mu\ra 3e$ experiments put the most stringent constraints on the model.
The existing  constraints already hint at  Yukawa couplings must
exhibit some special patterns in order that the model remains viable.
We gave two extreme examples to demonstrate how the Yukawa patterns help to
ameliorate the constraints from  experimental. Alternatively, if the main
features of the model were to be confirmed
 knowledge of Yukawa structure will be obtained from precise
flavor violating experiments.

Now we summarize the possible signatures of this model:
\begin{enumerate}

\item For very low energy, the neutrino-electron scattering will
receive corrections from all kinds of KK excitation. These
corrections can be as large as the SM quantum corrections but with
opposite trends. Neutrino flavor changing scattering clearly
indicate  new physics beyond SM but the rate is estimated to be too small for
detection.

\item With the linear colliders with $\sqrt{s}<1/R$, say $\sim 500$ GeV,
the Bhabha and M\o ller scattering spectrum can be very different  from SM.
The smoking gun evidence for bileptons  will be flavor changing
scattering, $e^-e^-\ra e^- \mu^-, \tau^-\mu^-, \cdots$ and
$e^+e^-\ra e^+\mu^-, \mu^+\mu^-, \cdots$. The rates depend on the
off-diagonal entities of ${\cal U}$ which cannot be predicted in
this model.

\item For  a  linear collider with $\sqrt{s}$ around $1/R$, there will be
strong resonance  enhancement for $e^-e^-\ra \mu^- \mu^-, \tau^-\tau^-$
and $e^+ e^-\ra e^+ \mu^-$ and so on. This are unmistakable
signals of vector bilepton.

\item With a multi-TeV linear collider that has energy reach of  $\sqrt{s}\sim 2/R>3$ TeV,
we can see the direct productions of gauge exotic bosons:
 $e^+ e^-\ra U^{\pm2}U^{\mp2}, V^\pm V^\mp$
 and  $e^+ e^-\ra W_0^\pm W_1^\mp$.
The extra $W_1 W_0$ channel will distinguish this 5D model
from  the 4D models with bileptons.
And at the same effective $\sqrt{s}$ range,
the hadron collider can produce single  $W_1^\pm$ KK mode but not
$V^\pm$ KK mode.
\end{enumerate}
Hopefully these can be seen in the next generation of experiments.

\acknowledgments{
Work was supported in part by the Natural Science and Engineering Council of
Canada.}

\appendix
\section{KK decomposition and KK number conservation law}
On the $S_1/(Z_2\times Z'_2)$ orbifold, the bulk fields can be
decomposed in terms of eigen-modes( $n>0$),
\beqa
\phi^{++}_n(y) &=& {1\over \sqrt{\pi R}} \cos \frac{2 n y}{R}\\
\phi^{+-}_n(y) &=& {1\over \sqrt{\pi R}} \cos \frac{(2 n-1) y}{R}\\
\phi^{-+}_n(y) &=& {1\over \sqrt{\pi R}} \sin \frac{(2 n-1) y}{R}\\
\phi^{--}_n(y) &=& {1\over \sqrt{\pi R}} \sin \frac{2 n y}{R}
\eeqa
and the only zero mode $\phi^{++}_0(y)= 1/\sqrt{2\pi R}$.
To simplify the notation, we will use
$\{e^n_1,o^n_1,o^n_2,e^n_2\}$ to denote the $n-$th modes of
$\phi^{++}$, $\phi^{+-}$, $\phi^{-+}$ and $\phi^{--}$
respectively.
The physical space is $y\in[0, \pi R/2]$. But the integration has
to be carried over the whole space $y\in[0,2\pi R]$ such
that they form a orthonormal basis:
\beqa
\langle e^n_1 | e^m_1 \rangle =\langle e^n_2 | e^m_2 \rangle =
\langle o^n_1 | o^m_1 \rangle =\langle o^n_2 | o^m_2 \rangle =
\delta_{m,n}\, ,\nonr\\
\langle e^n_1 | e^m_2 \rangle =\langle e^n_1 | o^m_1 \rangle =
\langle e^n_1 | o^m_2 \rangle =0\, ,\nonr\\
\langle e^n_2 | o^m_1 \rangle =
\langle e^n_2 | o^m_2 \rangle = \langle o^n_1 | o^m_2 \rangle
=0\, .\nonr
\eeqa
It can be found that
\beqa
 e_1^n e_1^m = {e_1^{n+m}+e_1^{n-m}\over 2},\,
 e_2^n e_2^m = {e_1^{n-m}-e_1^{n+m}\over 2},\nonr\\
 o_1^n o_1^m = {e_1^{n+m-1}+e_1^{n-m}\over 2},
 o_2^n o_2^m = {e_1^{n-m}-e_1^{n+m-1}\over 2},\nonr\\
 e_1^n e_2^m = {e_2^{n+m}-e_2^{n-m}\over 2},\,
 e_1^n o_1^m = {o_1^{n+m}+o_1^{n-m+1}\over 2},\nonr\\
 e_1^n o_2^m = {o_2^{n+m}-o_2^{n-m+1}\over 2},\,
 e_2^n o_1^m = {o_2^{n+m}+o_2^{n-m+1}\over 2},\nonr\\
 e_2^n o_2^m = {-o_1^{n+m}+o_1^{n-m+1}\over 2},
 o_1^n o_2^m = {e_2^{n+m-1}-e_2^{n-m}\over 2}.\nonr
\eeqa
Note that the KK number conservation rule now is
\beq
2n^e_1 \pm 2n^e_2 \pm \cdots \pm (2 n^o_1-1) \pm (2 n^o_2-1) \pm
\cdots =0
\eeq
where $n^e_i$($n^o_i$) is the KK number of the $i-$th even(odd) mode.
This is  different from the $S_1/Z_2$ case.
Taking 3 bulk fields (with KK number $l,m,n$) interaction
as an example, besides the usual $m\pm n\pm l=0 $ rule
there is another $ m \pm n \pm l=1$ rule for two odd-modes fuse
with one even-mode. It is easily  understood since the transformation
of $Z_2$ followed by $Z_2'$ is equal to the translation, $y\ra y+2\pi R$,
which will modify KK number by one.

\section{Some handy formulas }
Here we collect some useful identities for calculation.
The charge conjugation is defined as $\psi^\CC= C\bar{\psi}^T$ and
the matrix $C$ satisfies:
\beq
C^\dag C=1,\, C^T=-C,\, C \gamma^T_\mu C^{-1}=-\gamma_\mu
\eeq
 and also
\beq
C \sigma_{\mu\nu}^T C^{-1}= \sigma_{\mu\nu},\,
C \gamma_5^T C^{-1}=\gamma_5,\,
C (\gamma_\mu\gamma_5)C^{-1}=\gamma_\mu\gamma_5\,.
\eeq
We use the  representation : $C= -i \gamma^2\gamma^0$.
Also
$\bar{\psi^\CC}= - \psi^T C^{-1}$, and the helicity projections are
$P_{L/R}= (1\mp \gamma^5)/2$. The following relations can be derived
\beqa
\bar{\psi^\CC_1} P_{L/R} \psi_2=\bar{\psi^\CC_2} P_{L/R} \psi_1\,,\\
\bar{\psi_1} P_{L/R} \psi^\CC_2= +\bar{\psi_2} P_{L/R}
\psi_1^\CC\,,\\
\label{eq:llU}
\bar{\psi_1}\gamma^\mu P_{L/R}\psi^\CC_2 =
-\bar{\psi_2}\gamma^\mu P_{R/L} \psi^\CC_1\,,\\
\bar{\psi_1^\CC}\gamma^\mu P_{L/R}\psi_2 =
-\bar{\psi_2^\CC}\gamma^\mu P_{R/L} \psi_1\,,\\
\bar{\psi_1^\CC}P_{L/R}\psi_2^\CC = \bar{\psi_2}P_{L/R}\psi_1\,,\\
\bar{\psi_1^\CC} \gamma^\mu P_{L/R} \psi_2^\CC = - \bar{\psi_2}
\gamma^\mu P_{R/L}\psi_1\,,\\
(\bar{\psi_1}P_L \psi_3 )(\bar{\psi_2}P_R \psi_4)=\nonr\\
 -\frac12
(\bar{\psi_1}\gamma^\mu P_R\psi_4)(\bar{\psi_2} \gamma_\mu P_L
\psi_3)\,,\\
(\bar{\psi_1}\gamma^\mu P_{L/R} \psi_3 )(\bar{\psi_2}\gamma_\mu P_{L/R}
\psi_4) =\nonr\\
(\bar{\psi_1}\gamma^\mu P_{L/R} \psi_4 )(\bar{\psi_2}\gamma_\mu
P_{L/R}\psi_3)\,.
\eeqa

\section{Triple gauge coupling}

\begin{figure}[h]
\centering \includegraphics[width=0.9\columnwidth]{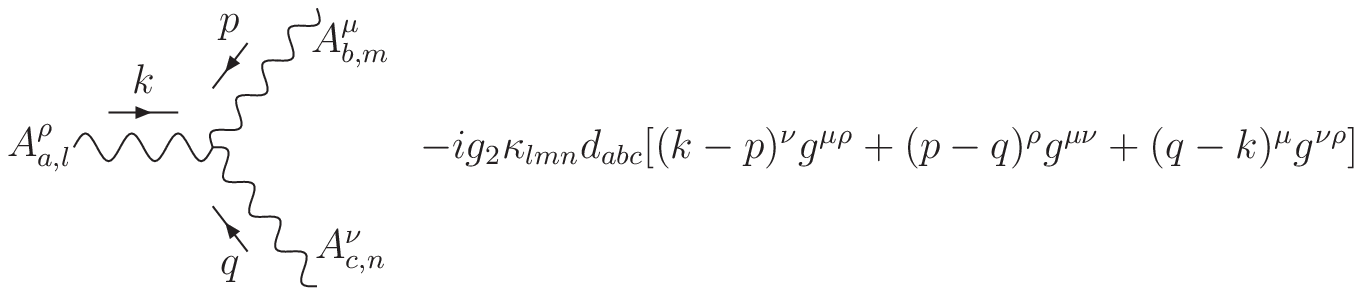}
\end{figure}
The Feynman rules for triple gauge boson coupling are summarized
in the following table.

\begin{center}
\begin{tabular}{|c|ccccccc|}
\hline
a & $A$ & $A$ & $A$ & $Z$ & $Z$ & $Z$ & $W^-$ \\
b & $W^+$ & $V^+$ & $U^{+2}$ & $W^+$ & $V^+$ & $U^{+2}$ & $U^{+2}$ \\
c & $W^-$ & $V^-$ & $U^{-2}$ & $W^-$ & $V^-$ & $U^{-2}$ & $V^{-}$ \\
\hline
$d_{abc}$ & $s_W$ & $s_W$ & $2s_W$& $c_W$ & $-{1\over 2c_W}$ & ${1-4s_W^2 \over
2c_W}$ & $\frac{1}{\sqrt{2}}$\\
\hline
\end{tabular}\end{center}

Where $a,b,c$ denote the  gauge boson species
and $d_{abc}$ can be easily determined by group structure.
The indices  $l,m,n$ collectively represent their corresponding KK numbers and $(Z_2,Z'_2)$
parity.
$\kappa$ is totally symmetric and is determined by
\beq
\kappa_{lmn}=\sqrt{2\pi R }\int^{2\pi R }_0 dy  \phi_l(y)
\phi_m(y) \phi_n(y).
\eeq
For example, $\kappa_{0_{++},n_{+-},n_{+-}}= \kappa_{0_{++},n_{++},n_{++}}=1$,
and   $\kappa_{2n_{++}, n_{++}, n_{++}}=1/\sqrt{2}$. In fact,
$\kappa=1$ when  any one of the three is a zero mode and all the
other allowed  combinations which respect KK number conservation and
$Z_2\times Z'_2$ parity give $\pm 1/\sqrt{2}$.

%\newpage

\end{document}